\documentclass[longauth]{aa}  

\usepackage{amsmath}
\usepackage{amssymb}
\usepackage{fixltx2e}
\usepackage[english]{babel}
\usepackage{graphicx}
\usepackage{epstopdf}
\usepackage{epsf,color}
\usepackage[mathscr]{eucal}
\usepackage{amsmath}
\usepackage{amssymb,amsfonts}
\usepackage{graphicx}
\usepackage{txfonts}
\usepackage{dsfont}
\definecolor{Mygreen}{rgb}{0.75, 0.0, 0.0}
\definecolor{Mypink}{rgb}{1.0, 0.0, 0.5}
\definecolor{Myred}{rgb}{0.7, 0.0, 0.0}
\usepackage[breaklinks, citecolor=blue, linkcolor=Myred, urlcolor=Myred, colorlinks=true]{hyperref}
\usepackage{float} 
\usepackage{color}
\usepackage{ulem}
\usepackage{hyperref}
\usepackage{mwe,tikz}
\usepackage[percent]{overpic}
\usepackage{booktabs}

\usepackage{natbib}
\bibpunct{(}{)}{;}{a}{}{,}

\begin{document}

\title{The XXL Survey}
\subtitle{LI. Pressure profile and $Y_{\rm SZ} - M$ scaling relation in three low-mass galaxy clusters at $z\sim1$ observed with NIKA2\thanks{The FITS file of the published NIKA2 maps are only available at the CDS
via anonymous ftp to cdsarc.u-strasbg.fr (XXXXX) or via XXXXX}}

\author{
R.~Adam \inst{\ref{OCA},\ref{LLR}}
\and M.~Ricci \inst{\ref{APC},\ref{LAPP},\ref{OCA}}
\and D.~Eckert \inst{\ref{geneva}}
\and  P.~Ade \inst{\ref{Cardiff}}
\and  H.~Ajeddig \inst{\ref{CEA}}
\and  B.~Altieri\inst{\ref{ESAC}}
\and  P.~Andr\'e \inst{\ref{CEA}}
\and  E.~Artis \inst{\ref{LPSC}}
\and  H.~Aussel \inst{\ref{CEA}}
\and  A.~Beelen \inst{\ref{LAM}}
\and  C.~Benoist\inst{\ref{OCA}}
\and  A.~Beno\^it \inst{\ref{Neel}}
\and  S.~Berta \inst{\ref{IRAMF}}
\and  L.~Bing \inst{\ref{LAM}}
\and  M.~Birkinshaw\inst{\ref{Bristol}}\thanks{Deceased}
\and  O.~Bourrion \inst{\ref{LPSC}}
\and  D.~Boutigny \inst{\ref{LAPP}}
\and  M.~Bremer\inst{\ref{Bristol}}
\and  M.~Calvo \inst{\ref{Neel}}
\and  A.~Cappi\inst{\ref{OCA},\ref{BolognaObs}}
\and  A.~Catalano \inst{\ref{LPSC}}
\and  M.~De~Petris\inst{\ref{Roma}}
\and  F.-X.~D\'esert \inst{\ref{IPAG}}
\and  S.~Doyle \inst{\ref{Cardiff}}
\and  E.~F.~C.~Driessen \inst{\ref{IRAMF}}
\and  L.~Faccioli\inst{\ref{CEA}}
\and  C.~Ferrari\inst{\ref{OCA}}
\and  F.~Gastaldello\inst{\ref{Milan}}
\and  P.~Giles\inst{\ref{Sussex}}
\and  A.~Gomez \inst{\ref{CAB}}
\and  J.~Goupy \inst{\ref{Neel}}
\and  O.~Hahn\inst{\ref{OCA}}
\and  C.~Hanser \inst{\ref{LPSC}}
\and  C.~Horellou\inst{\ref{Onsala}}
\and  F.~K\'eruzor\'e \inst{\ref{Argonne}}
\and  E.~Koulouridis\inst{\ref{Athene},\ref{CEA}}
\and  C.~Kramer \inst{\ref{IRAMF}}
\and  B.~Ladjelate \inst{\ref{IRAME}}
\and  G.~Lagache \inst{\ref{LAM}}
\and  S.~Leclercq \inst{\ref{IRAMF}}
\and  J.-F.~Lestrade \inst{\ref{LERMA}}
\and  J.F.~Mac\'ias-P\'erez \inst{\ref{LPSC}}
\and  S.~Madden \inst{\ref{CEA}}
\and  B.~Maughan\inst{\ref{Bristol}}
\and  S.~Maurogordato\inst{\ref{OCA}}
\and  A.~Maury \inst{\ref{CEA}}
\and  P.~Mauskopf \inst{\ref{Cardiff},\ref{Arizona}}
\and  A.~Monfardini \inst{\ref{Neel}}
\and  M.~Mu\~noz-Echeverr\'ia \inst{\ref{LPSC}}
\and  F.~Pacaud\inst{\ref{Bonn}}
\and  L.~Perotto \inst{\ref{LPSC}}
\and  M.~Pierre\inst{\ref{CEA}}
\and  G.~Pisano \inst{\ref{Roma}}
\and  E.~Pompei\inst{\ref{ESO}}
\and  N.~Ponthieu \inst{\ref{IPAG}}
\and  V.~Rev\'eret \inst{\ref{CEA}}
\and  A.~Rigby \inst{\ref{Cardiff}}
\and  A.~Ritacco \inst{\ref{INAF},\ref{ENS}}
\and  C.~Romero \inst{\ref{UPenn}}
\and  H.~Roussel \inst{\ref{IAP}}
\and  F.~Ruppin \inst{\ref{IP2I}}
\and  M.~Sereno\inst{\ref{BolognaObs},\ref{BolognaDepP}}
\and  K.~Schuster \inst{\ref{IRAMF}}
\and  A.~Sievers \inst{\ref{IRAME}}
\and  G.~Tintor\'e Vidal \inst{\ref{LLR}}
\and  C.~Tucker \inst{\ref{Cardiff}}
\and  R.~Zylka \inst{\ref{IRAMF}}
}

\offprints{Rémi Adam (\url{remi.adam@oca.eu})}

\institute{
Laboratoire Lagrange, Universit\'e C\^ote d'Azur, Observatoire de la C\^ote d'Azur, CNRS, Blvd de l'Observatoire, CS 34229, 06304 Nice cedex 4, France
  \label{OCA}
  \and
 LLR, CNRS, \'Ecole Polytechnique, Institut Polytechnique de Paris
  \label{LLR}
  \and
  Universit\'e Paris Cit\'e, CNRS, Astroparticule et Cosmologie, F-75013 Paris, France
  \label{APC}
  \and
Laboratoire d'Annecy de Physique des Particules, Universit\'e Savoie Mont Blanc, CNRS/IN2P3, F-74941 Annecy, France
  \label{LAPP}
  \and
  Department of Astronomy, University of Geneva, ch. d'Ecogia 16, CH-1290 Versoix, Switzerland
   \label{geneva}
   \and
    School of Physics and Astronomy, Cardiff University, Queen’s Buildings, The Parade, Cardiff CF24 3AA, UK
  \label{Cardiff}
  \and
Universit\'e Paris-Saclay, Universit\'e Paris Cit\'e, CEA, CNRS, AIM, F-91191, Gif-sur-Yvette, France (2022)
  \label{CEA}
  \and
European Space Astronomy Centre (ESA/ESAC), Operations Department, Villanueva de la Can\~{a}da, Madrid, Spain
\label{ESAC}
  \and
  Laboratoire de Physique Subatomique et de Cosmologie, Universit\'e Grenoble Alpes, CNRS/IN2P3, 53, avenue des Martyrs, Grenoble, France
  \label{LPSC}
\and
Aix Marseille Universit\'e, CNRS, LAM (Laboratoire d'Astrophysique de Marseille) UMR 7326, 13388, Marseille, France
  \label{LAM}
  \and
Institut N\'eel, CNRS and Universit\'e Grenoble Alpes, France
  \label{Neel}
    \and
Institut de RadioAstronomie Millim\'etrique (IRAM), Grenoble, France
  \label{IRAMF}
  \and
HH Wills Physics Laboratory, University of Bristol, Tyndall Avenue, Bristol, BS8 1TL, UK
\label{Bristol} 
\and
  Istituto Nazionale di Astrofisica (INAF) - Osservatorio di Astrofisica e Scienza dello Spazio (OAS), via Gobetti 93/3, I-40127 Bologna, Italy
\label{BolognaObs}
\and
Dipartimento di Fisica, Sapienza Universit\`a di Roma, Piazzale Aldo Moro 5, I-00185 Roma, Italy
  \label{Roma}
\and
Univ. Grenoble Alpes, CNRS, IPAG, F-38000 Grenoble, France 
  \label{IPAG}
 \and
INAF - IASF Milan, via A. Corti 12, I-20133 Milano,  Italy
\label{Milan}  
\and
Department of Physics and Astronomy, University of Sussex, Brighton BN1 9QH, UK
\label{Sussex} 
\and
Centro de Astrobiolog\'ia (CSIC-INTA), Torrej\'on de Ardoz, 28850 Madrid, Spain
\label{CAB}
\and
Department of Space, Earth and Environment, Chalmers University of Technology, Onsala Space Observatory, SE-439 92 Onsala,
Sweden
\label{Onsala}
\and
High Energy Physics Division, Argonne National Laboratory, 9700 South Cass Avenue, Lemont, IL 60439, USA
\label{Argonne}
\and
Institute for Astronomy \& Astrophysics, Space Applications \& Remote
Sensing, National Observatory of Athens, GR-15236 Palaia Penteli,Greece
  \label{Athene}
  \and
Institut de RadioAstronomie Millim\'etrique (IRAM), Granada, Spain 
\label{IRAME}
\and 
LERMA, Observatoire de Paris, PSL Research University, CNRS, Sorbonne Universités, UPMC Univ. Paris 06, 75014 Paris, France
  \label{LERMA}  
\and
School of Earth and Space Exploration and Department of Physics, Arizona State University, Tempe, AZ 85287
  \label{Arizona}
\and 
Argelander Institut f\"ur Astronomie, Universit\"at Bonn, Auf dem Huegel 71, DE-53121 Bonn, Germany
\label{Bonn}
\and
European Southern Observatory, Alonso de Cordova 3107, Vitacura,
19001 Casilla, Santiago 19, Chile
\label{ESO}
\and
INAF-Osservatorio Astronomico di Cagliari, Via della Scienza 5, 09047 Selargius, Italy
\label{INAF}
\and
LPENS, Ecole Normale Sup\'erieure, 24 rue Lhomond, 75005, Paris (FR) 
\label{ENS}
\and
Department of Physics and Astronomy, University of Pennsylvania, 209 South 33rd Street, Philadelphia, PA 19104, USA
\label{UPenn}
\and 
Institut d'Astrophysique de Paris, Sorbonne Universit\'es, UPMC Univ. Paris 06, CNRS UMR 7095, 75014 Paris, France 
  \label{IAP}
\and
Univ. Lyon, Univ. Claude Bernard Lyon 1, CNRS/IN2P3, IP2I Lyon, 69622 Villeurbanne, France
\label{IP2I}
\and
INFN, Sezione di Bologna, viale Berti Pichat 6/2, 40127 Bologna, Italy
\label{BolognaDepP}
}

\date{Received \today \ / Accepted --}
\abstract 
{The thermodynamical properties of the intracluster medium (ICM) are driven by scale-free gravitational collapse, but they also reflect the rich astrophysical processes at play in galaxy clusters. At low masses ($\sim 10^{14}$ M$_{\odot}$) and high redshift ($z \gtrsim 1$), these properties remain poorly constrained, observationally speaking, due to the difficulty in obtaining resolved and sensitive data.}
{We aim to investigate the inner structure of the ICM as seen through the Sunyaev-Zel'dovich (SZ) effect in this regime of mass and redshift. We focused on the thermal pressure profile and the scaling relation between SZ flux and mass, namely the $Y_{\rm SZ} - M$ scaling relation.}
{The three galaxy clusters XLSSC~072 ($z=1.002$), XLSSC~100 ($z=0.915$), and XLSSC~102 ($z=0.969$), with $M_{500} \sim 2 \times 10^{14}$ M$_{\odot}$, were selected from the XXL X-ray survey and observed with the NIKA2 millimeter camera to image their SZ signal. XMM-Newton X-ray data were used as a complement to the NIKA2 data to derive masses based on the $Y_X - M$ relation and the hydrostatic equilibrium.}
{The SZ images of the three clusters, along with the X-ray and optical data, indicate dynamical activity related to merging events. The pressure profile is consistent with that expected for morphologically disturbed systems, with a relatively flat core and a shallow outer slope. Despite significant disturbances in the ICM, the three high-redshift low-mass clusters follow the $Y_{\rm SZ}-M$ relation expected from standard evolution remarkably 
well.}
{These results indicate that the dominant physics that drives cluster evolution is already in place by  $z \sim 1$, at least for systems with masses above $M_{500} \sim 10^{14}$ M$_{\odot}$.}

\titlerunning{The XXL Survey LV}
\authorrunning{R. Adam et al.}
\maketitle

\section{Introduction}
\label{sec:introduction}
The presence of a diffuse hot gas component that permeates galaxy clusters, the intracluster medium (ICM), was revealed by X-ray observations in the 1970s (see \citealt{Sarazin1986} and \citealt{Biviano2000} for historical reviews). Thanks to subsequent observational and theoretical achievements, it is now established that clusters are dominated by dark matter ($\sim 80$\%), that the ICM accounts for most of their baryonic matter ($\sim 12$\%), and that galaxies only provide the remaining few percent of the total cluster mass. Clusters form at the intersection of cosmic filaments and trace the growth of cosmic structures, as peaks in the matter density field. As such, they are recognized as unique astrophysical laboratories and as important cosmological probes \citep[e.g.,][]{Vikhlinin2009a,PlanckXX2014,Bocquet2019}. We invite the reader to consult \cite{Allen2011} for a review.

The formation of galaxy clusters and their ICM is mainly driven by the gravitational collapse of large-scale structures \citep{Kravtsov2012}. This implies that, to first order, clusters are self-similar objects \citep{Kaiser1986} whose observational properties follow well-predicted behaviors once rescaled in mass and redshift. However, clusters are also affected by complex astrophysical processes related to gas dynamics and the formation of galaxies. The ICM is believed to be established in the early phase of cluster formation history and to be continuously fed by the merging of smaller groups and the accretion of the surrounding material; the infalling gas kinetic energy is mostly converted into heat via shocks and turbulent cascades. In parallel, a fraction of the baryons condensates into stars, eventually inducing significant supernovae or active galactic nucleus (AGN) feedback onto the ICM \citep{Fabian2012,Hlavacek2022}. These processes should leave an imprint both in the inner structure of the ICM and in the scaling relation between global integrated properties of the clusters \citep{Lovisari2022}. Characterizing the ICM thermodynamics is therefore an excellent way to address the nature of the astrophysical processes associated with cluster formation; it is a necessary step when using clusters to probe the growth of large-scale structures \citep{Voit2005}.

The thermal Sunyaev-Zel'dovich (SZ) effect \citep{Sunyaev1970,Sunyaev1972} provides an independent and complementary probe of the ICM to X-ray observations \citep[][for reviews]{Birkinshaw1999,Mroczkowski2019}. It is due to the inverse Compton scattering of cosmic microwave background (CMB) photons on ICM electrons. Its surface brightness is independent of redshift, which makes it particularly competitive for distant objects provided that sufficiently high angular resolution and sensitivity are available (e.g., \citealt{Korngut2011}, \citealt{Kitayama2016}, and \citealt{Adam2016} for the use of relevant facilities). Unlike X-ray observations, which rely on the combination of gas density and temperature (measured using spectroscopy) to infer the pressure, the SZ effect directly measures the thermal ICM electron pressure. The pressure profile is an excellent tracer of the matter distribution because it reflects how the gas is compressed in the potential well. Similarly, the SZ flux, $Y_{\rm SZ}$, tracks the total mass with a low intrinsic scatter since it measures the thermal energy directly, itself related to the depth of the potential well \citep{Nagai2007,Pratt2019}. Therefore, the SZ effect is also an excellent way to detect clusters, with a clean selection function, and it has been given much attention not only for studying the SZ signal directly, but also in follow-up observations of SZ-selected samples \citep[e.g.,][]{Sanders2018,Bartalucci2019}. For all these reasons, the $Y_{\rm SZ}-M$ relation and the pressure profile are key tools that require detailed characterization if one wants to fully benefit from the statistical power of SZ surveys for cluster cosmology and astrophysics. 

The thermal pressure profile and the $Y_{\rm SZ}-M$ scaling relation have been deeply investigated and calibrated up to intermediate redshifts \citep[e.g.,][]{Arnaud2010,PlanckV2013,Ghirardini2019,Bonamente2008,PlanckXX2014,Medezinski2018}. However, nontrivial redshift evolution is expected \citep{LeBrun2017} because of the changes in the cluster mass-accretion rate and the merger activity \citep{Fakhouri2010,Fakhouri2010b} or the enhanced star formation and AGN activity \citep{Alberts2016}. Yet, current attempts, either in SZ or in follow-up X-ray observations of SZ-selected samples, did not report significant nonstandard evolution of the bulk ICM properties for massive systems out to $z \sim 2$ (see \citealt{McDonald2017}, \citealt{Bartalucci2017}, \citealt{Mantz2018} (hereafter XXL Paper XVII), and \citealt{Ghirardini2021}). At lower masses, however, clusters are expected to be more affected by the gas dynamics and the interaction with galaxies due to their shallower potential well, so supplementary deviations in the ICM properties are expected in this regime \citep[$M \lesssim 10^{14}$ M$_{\odot}$, e.g.,][]{Pop2022}. Although low-mass clusters at high redshifts will represent a large fraction of the detections of ongoing and future cluster surveys (e.g., eROSITA, Euclid, LSST, CMB-Stage4, see \citealt{Pillepich2012}, \citealt{Bulbul2022}, \citealt{EuclidIII2019}, \citealt{LSST2009}, and \citealt{Abazajian2016}), their detailed SZ observational properties remain nearly unexplored to date due to the difficulty in obtaining sufficiently high-quality data \citep[see][for MUSTANG2 observations]{Dicker2020}. Given their expected high sensitivity to cluster astrophysics, dedicated SZ follow-ups with resolved observations are thus becoming essential in efforts to further advance cluster cosmology and astrophysics.

In this paper, we report on SZ observations of three low-mass ($M_{500} \sim 2 \times 10^{14}$ M$_{\odot}$\footnote{$M_{500}$ is the mass enclosed within $R_{500}$, the radius within which the mean cluster density is 500 times the critical density of the Universe at the given redshift.}) high-redshift ($z \sim 1$) galaxy clusters selected from the XXL X-ray survey \citep[][hereafter XXL Paper I]{Pierre2016}: XLSSC~072 ($z=1.002$), XLSSC~100 ($z=0.915$), and XLSSC~102 ($z=0.969$). They were imaged with the New IRAM KIDs Array 2 (NIKA2) millimeter camera \citep{Adam2018a} from the Institut de Radio Astronomie Millimétrique (IRAM) 30-meter telescope at 150 GHz and 260 GHz. Given the nature of these objects in terms of mass and redshift, we aim to test the standard evolution of the ICM as calibrated on nearby massive systems in a regime that has not been explored with resolved SZ data yet and where astrophysical processes should be more effective. The data, complemented with X-ray and optical observations are used to investigate the dynamical state of the clusters. The SZ images are used to derive the thermal pressure profiles and extract the SZ fluxes, which are compared with standard evolution expectations once the masses are extracted under different assumptions using SZ and/or X-ray data. This work extends over the earlier multiwavelength analysis of XLSSC~102 \citep{Ricci2020}, hereafter XXL Paper XLIV\footnote{In Paper XLIV, optical, X-ray and NIKA2 SZ data were used to analyze XLSSC~102. It was found that the cluster experienced a major merging event, which shifted the positions of gas and galaxies. The thermodynamic profiles of the cluster were measured, indicating characteristics typical of disturbed systems. The impact of local pressure substructure and the cluster center definition was investigated, and the global properties of XLSSC~102 were compared to other high-mass, low-redshift clusters. No strong evidence of unusual evolution was observed.}, by adding new NIKA2 observations for two clusters, the use of new X-ray data for XLSSC~102, and by extending the analysis methodology to recover the cluster pressure profile and their masses.

The paper is organized as follows. In Section \ref{sec:data}, we present the different datasets used in this work. In Section \ref{sec:dynamics}, we describe the multiwavelength morphological comparison and discuss the dynamical state of the clusters. We present the modeling and the data analysis methodology in Section \ref{sec:modeling}. The results are reviewed in Section \ref{sec:results}, and the main conclusions are summarized in Section \ref{sec:Summary_and_conclusions}. Throughout the paper, we assume a flat $\Lambda$CDM cosmology with $H_0 = 70$ km s$^{-1}$ Mpc$^{-1}$ and $\Omega_{\rm m} = 0.3$. The Hubble parameter at redshift $z$ normalized to the present-day value is defined as $E(z) = H(z)/H_0$.

\section{Data}\label{sec:data}
In this section, we discuss the selection of the clusters and present the main data that were used.

\subsection{Cluster sample}
\begin{table*}[h]
\caption{\footnotesize{Summary of the survey properties of the XXL targeted clusters. The masses obtained according to the mass-temperature relation are labeled $M_{500, \rm MT}$. The value of $M_{500, \rm MT}$ based on Paper XX was computed from $R_{500}$ accounting for the different cosmological model.}}
\begin{center}
\resizebox{\textwidth}{!} {
\begin{tabular}{c|ccc|cc|cccc|cccc}
\hline
\hline
ID & R.A.$^{(a)}$ & Dec.$^{(a)}$ & $z^{(a)}$ & $T_{300 \rm \ kpc}^{(a)}$ & $T_{300 \rm \ kpc}^{(b)}$ & $M_{500, \rm MT}^{(a)}$ & $M_{500, \rm MT}^{(c)}$ & $M_{500, \rm MT}^{(d)}$ & $M_{500, \rm scal}^{(a)}$ & $M_{500, \rm UPP}^{(e)}$ & $M_{500, \rm Cal}^{(e)}$ & $M_{500, \rm UPP}^{(f)}$ & $M_{500, \rm Cal}^{(f)}$ \\
 (---) & (degree) & (degree) & (---) & (keV) & (keV) & ($10^{14}$ M$_{\odot}$) & ($10^{14}$ M$_{\odot}$) & ($10^{14}$ M$_{\odot}$) & ($10^{14}$ M$_{\odot}$) & ($10^{14}$ M$_{\odot}$) & ($10^{14}$ M$_{\odot}$) & ($10^{14}$ M$_{\odot}$) & ($10^{14}$ M$_{\odot}$)\\
\hline
XLSSC 072 & $33.850$ & $-3.726$ & $1.002$ & $2.00^{+0.27}_{-0.31}$ & $3.7_{-0.6}^{+1.1}$ & 0.70 & $1.9\pm1.1$ & $0.69^{+0.69}_{-0.35}$ & $2.58 \pm 1.08$ & $2.46^{+0.44}_{-0.37}$ & $3.61^{+0.87}_{-0.74}$ & N.A. & N.A. \\[0.1cm]
XLSSC 100 & $31.549$ & $-6.193$ & $0.915$ & $5.60^{+0.51}_{-0.43}$ & $4.3_{-1.2}^{+1.7}$ & 4.08 & $2.6\pm1.8$ & $1.78^{+1.73}_{-0.92}$ & $2.55 \pm 1.08$ & N.A. & N.A. & N.A. & N.A. \\ [0.1cm]
XLSSC 102 & $31.322$ & $-4.652$ & $0.969$ & $3.87^{+0.81}_{-0.76}$ & $3.2_{-0.5}^{+0.8}$ & 2.13 & $1.9\pm1.1$ & $1.17^{+1.16}_{-0.60}$ & $2.64 \pm 1.09$ & $3.12^{+0.52}_{-0.44}$ & $4.59^{+1.06}_{-0.99}$ & $3.16^{+0.51}_{-0.44}$ & $4.44^{+0.85}_{-0.76}$ \\ 
\hline
\end{tabular} 
}
\end{center}
{\footnotesize {\bf Notes.} 
$^{(a)}$Paper XX.
$^{(b)}$Paper III.
$^{(c)}$Paper II.
$^{(d)}$\cite{Umetsu2020}.
$^{(e)}$\cite{Hilton2018}.
$^{(f)}$\cite{Hilton2021}.
}
\label{tab:sample_summary}
\end{table*}

The target clusters were selected from the XXL survey (see also \citealt{Ricci2018PhD} for details), performed with XMM-Newton in the X-rays (Paper I). Thanks to its selection function, the XXL survey allowed us to identify clusters at low masses and high redshift \citep[][hereafter XXL Paper II]{Pacaud2016}. We only considered the most securely detected XXL clusters (classified as C1) from the northern part of the XXL survey (XXL-N) that are observable from the IRAM 30m telescope. We requested detections in the optical using the galaxy overdensity to make sure that the clusters were also confirmed independently from the ICM content (see Section~\ref{sec:optical_data} for details). Only detections with robust spectroscopic redshift estimates were accounted for (Paper XX). According to these criteria, we selected the three clusters at redshift $z \sim 1$ with X-ray data of sufficient quality to allow a reliable combination with NIKA2, both in terms of images and radial profiles. Although the number of objects is limited, we expect them to be representative of other systems with similar parameters. All targets are part of the 100 brightest XXL clusters (Paper II): XLSSC~072, XLSSC~100, and XLSSC~102.

Different mass estimates were obtained from the XXL survey. In Paper XX, count rates together with scaling relations were used iteratively to infer the mass and the temperature without relying on X-ray spectroscopy. Spectroscopic temperatures were extracted within 300 kpc of the cluster center from XMM-Newton \citep[][hereafter XXL Paper III and XXL Paper XX]{Giles2016,Adami2018}. The masses were then estimated according to the mass-temperature relation calibrated using weak lensing data \citep[][XXL Paper IV]{Lieu2016}. A similar approach was also performed by \cite{Umetsu2020}, using Paper XX temperatures, where more reliable weak lensing masses were measured thanks to Hyper Suprime-Cam data \citep{Aihara2022} and to the use of a less restrictive prior on the concentration-mass relation.

The three selected clusters are located in the footprint surveyed by the Atacama Cosmology Telescope \citep[ACT,][]{Hilton2018,Hilton2021}. The ACT cluster catalog reports detections for which the signal-to-noise ratio (S/N) is larger than four. The masses were obtained by matching the normalization of the universal pressure profile (UPP) as calibrated by \cite{Arnaud2010} to the ACT data. Masses rescaled using a richness-based weak-lensing mass calibration factors are also provided. XLSSC~102 is reported in both the \cite{Hilton2018} and \cite{Hilton2021} catalogs with a S/N of 8.3 and 12.1, respectively. XLSSC~100 is not reported in either catalog. XLSSC~072 is only reported in \cite{Hilton2018}, with a S/N of 6.5.

In the case of XLSSC~072, a dedicated analysis was performed by \cite{Duffy2022}, hereafter XXL Paper XLVIII, using a deep XMM-Newton follow-up of the target, and thus clearly of higher quality than the survey data used in the previous XXL analysis. They obtained $T_{\rm 300 \ kpc} = 4.9^{+0.8}_{-0.6}$ keV, $T_{R_{500}} = 4.5 \pm 0.6$ keV and an hydrostatic mass $M_{500} = 2.4^{+1.7}_{-0.8} \times 10^{14}$ M$_{\odot}$ thanks to the NFW mass model fit to the data and a temperature profile resolved in four bins. We note that the temperature estimate of Paper XX is surprisingly much lower than that of Paper XLVIII, the later being more reliable given the better data quality. The mass estimate of \cite{Umetsu2020}, which used Paper XX temperatures, is consequently lower than what would have been obtained with alternative temperature measurements.

Our targets have masses $M_{500} \sim 2\times10^{14}$ M$_{\sun}$ according to XXL estimates. This is in rough agreement with the mass of XLSSC~102 as already measured in Paper XLIV, $M_{500} \sim \left(1 - 2\right) \times 10^{14}$ M$_{\odot}$, depending on the analysis choices (see Table~4 of Paper XLIV). We note that the ACT masses are significantly larger than those from XXL, depending on the assumptions, especially for XLSSC~102. The main properties of the cluster sample are summarized in Table~\ref{tab:sample_summary}.

\subsection{NIKA2}
\begin{table*}[h]
\caption{\footnotesize{Observational summary of the NIKA2 observations after data selection.}}
\begin{center}
\resizebox{\textwidth}{!} {
\begin{tabular}{c|cccccccc}
\hline
\hline
ID & Pointing center & $N_{\rm scan}$ & $t_{\rm obs}$ & Opacity & Beam FWHM & Absolute calibration accuracy & Pointing accuracy & Peak S/N$^{\dagger}$ \\
-- & R.A, Dec (deg) & -- & hours & 150, 260 GHz & 150, 260 GHz (arcsec) & 150, 260 GHz (\%) & 150, 260 GHz (arcsec) & -- \\
\hline
XLSSC~072 & $+33.850$, $-3.726$ & $123$ & $10.0$ & $0.19$, $0.32$ & $17.8$, $12.4$ & $3.6$, $14.8$ & $2.9$, $2.9$ & $-9.7$ \\ 
XLSSC~100 & $+31.549$, $-6.193$ & $122$ & $10.0$ & $0.19$, $0.32$ & $17.9$, $12.3$ & $5.6$, $18.0$ & $1.5$, $1.8$ & $-9.2$ \\  
XLSSC~102 & $+31.322$, $-4.652$ & $83$  &  $6.6$ & $0.15$, $0.24$ & $18.0$, $12.1$ & $3.8$, $7.7$  & $2.2$, $2.3$ & $-6.9$ \\   
\hline
\end{tabular}
}
\end{center}
{\footnotesize {\bf Notes.} 
$^{\dagger}$Peak S/N at an effective resolution of 27 arcsec FWHM at 150 GHz.
}
\label{tab:obs_summary}
\end{table*}

The clusters XLSSC~072, XLSSC~100, and XLSSC~102 were observed from January 2018 to February 2020 with the NIKA2 camera \citep{Adam2018a} at the IRAM 30m telescope, under projects 179-17, 094-18, 208-18, 093-19, 218-19, and 076-20. The observation scheme and the data reduction are similar for the three targets. We refer the reader to Paper XLIV for further details, where the XLSSC~102 data at 150 GHz were already presented. 

In brief, the beam was monitored using Uranus observations. Pointing corrections were checked using nearby quasars about every hour. The observing conditions were overall stable with average zenith opacity for the period \citep[for the methodology, see][]{Catalano2014}. The absolute calibration uncertainty was estimated using the dispersion of the flux measured from the observations of Uranus that bracket the clusters observations. Table~\ref{tab:obs_summary} summarizes the observational details for all three clusters, which are in line with the characteristics of the instrument given in \cite{Perotto2020}.

In the case of XLSSC~072, part of the data (October 2018; 25\% of the total observing time) could not be used with the standard calibration procedure due to a failure in the software control. For these data, the 150 GHz channel calibration was performed using the sufficiently bright radio source FIRST~J021511.4-034309 (or XXL-GMRT~J021511.4-034309, $\sim 30$ mJy at 150 GHz), located about 3 arcmin west of the cluster X-ray peak, assuming a constant flux as measured over the rest of the observing time. At 260 GHz, the source was too faint for proper measurement and the data could not be used. The resulting absolute calibration accuracy at 150 GHz was estimated to be 30\% for this subset, and the pointing accuracy was verified to be within a few arcsec using the dispersion of fluxes over the reliable scans used for cross-calibration. More information on the data validation can be found in Appendix~\ref{app:xlssc072_calib}.

The data were reduced as described in \cite{Adam2015}, by combining the individual detector time lines to remove the contribution from the electronic and atmospheric noise. The individual scan maps were checked and flagged based on the presence of large correlated noise residuals, prior to co-adding them using inverse variance weighting. The astrophysical signal filtering induced by the data reduction was estimated by injecting a fake signal in the data and comparing the input and output as a function of angular scale (the cluster signal is filtered at scales larger than the field of view of about 6.5 arcmin; see \citealt{Adam2015} for details). The statistical properties of the noise were derived by computing the power spectrum of half-difference maps obtained by dividing the dataset into equal parts. Monte Carlo (MC) noise realizations were computed using this power spectrum and preserving the noise standard deviation as a function of coordinates, following \cite{Adam2016}.

\begin{figure*}
        \centering
        \includegraphics[width=0.95\textwidth]{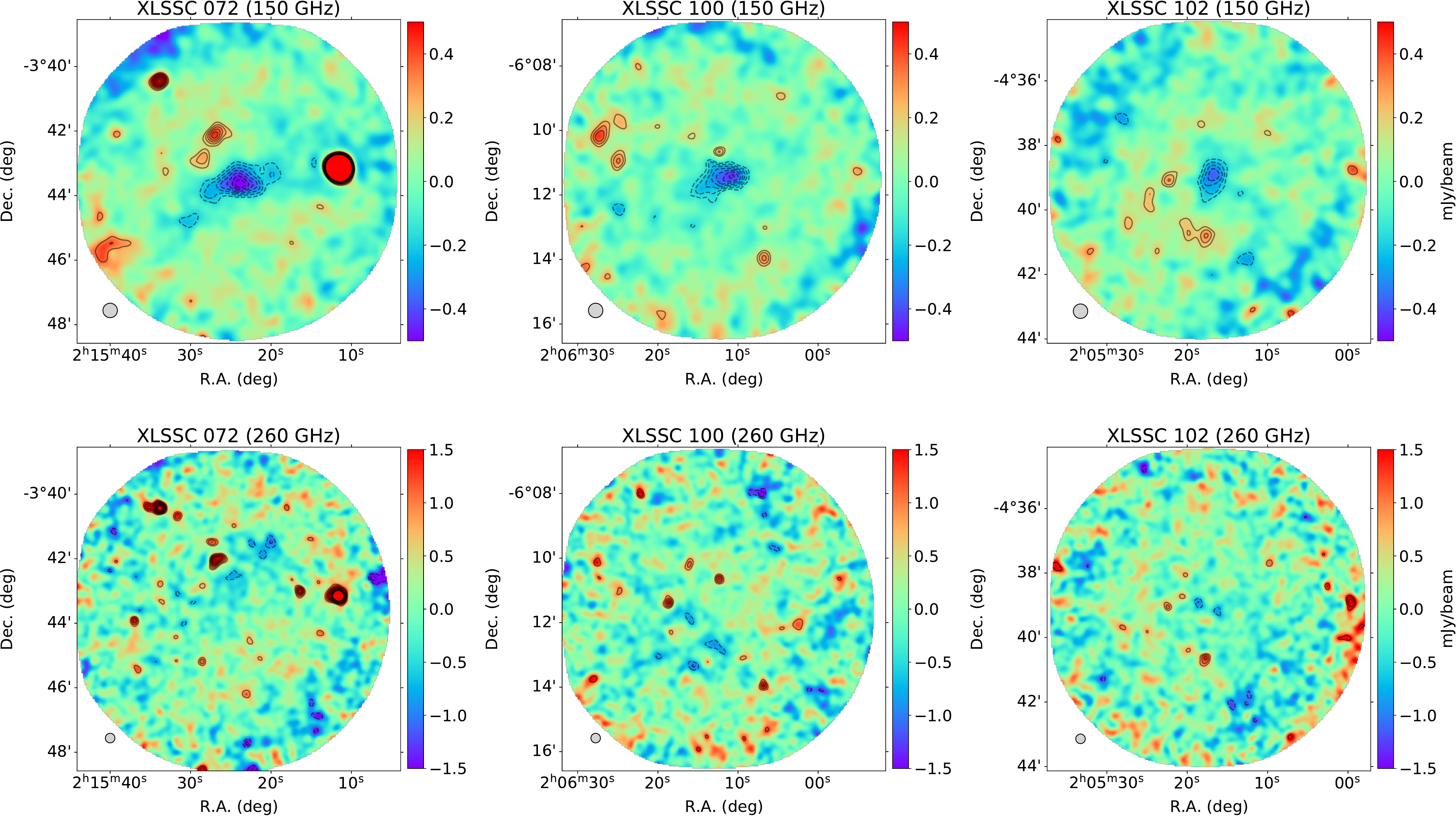}
        \caption{NIKA2 images at 150 GHz (top) and 260 GHz (bottom) for XLSSC~072, XLSSC~100, and XLSSC~102, from left to right. The maps have been smoothed to an effective resolution of 18 and 27 arcsec at 260 and 150 GHz, respectively. S/N contours are shown with 1$\sigma$ spacing starting at $\pm 3\sigma$. Data where the noise is greater than three times the value of the noise at the center of the map are masked. The effective beam size is shown in the bottom left corner of each panel.}
\label{fig:nika2maps}
\end{figure*}

In Figure~\ref{fig:nika2maps}, we present the NIKA2 surface brightness images of the targets at 150 and 260 GHz. All three clusters are well detected and show an extended decrement at 150 GHz, as expected for the SZ effect. At an effective resolution of 27 arcsec, the peak S/N is $-9.7$, $-9.2$, and $-6.9$, for XLSSC~072, XLSSC~100, and XLSSC~102, respectively. Given the 150 GHz data, the SZ signal is expected to peak at about 0.1 mJy/beam at 260 GHz and thus be well below the noise level. Several infrared and radio galaxies are also visible at both frequencies. They were identified and subtracted in the rest of the study, as discussed in Appendix~\ref{app:point_sources}, and we do not expect their contamination to play a significant role in the presented work.

\subsection{XMM-Newton}
We analyzed the \textit{XMM-Newton} data around the position of the three clusters of interest using the \textit{XMM-Newton} Science Analysis Software (\textsc{XMMSAS}) v16.1. We used the pipeline developed for the XMM-Newton Cluster Outskirts Project \citep[X-COP,][]{Eckert2017} to analyze the data. Namely, we applied the standard event selection criteria by running the XMMSAS tasks \texttt{emchain} and \texttt{epchain}. We then filtered out regions of enhanced soft proton background using the \texttt{mos-filter} and \texttt{pn-filter} executables to create clean event lists. Then we extracted X-ray photon maps for each observation in the $[0.5-2]$ keV bands by selecting all the valid events in the energy band of interest. We used the \texttt{eexpmap} task to extract effective exposure maps, which allowed us to take the telescope's vignetting into account. Finally, we used the \texttt{mos-spectra} and \texttt{pn-spectra} executables to create maps of the non-X-ray background by rescaling filter-wheel-closed data to the count rates measured in the unexposed corners of the telescope. Next, we stacked the extracted count maps, exposure maps and background maps for the three detectors (MOS1, MOS2, and PN). For more details on the data reduction technique, we refer the reader to \citet{Ghirardini2019}. 

We applied the aforementioned processing to all the observations of the survey and then co-added all the observations to create mosaic images of the entire XXL area. From the resulting mosaic images, we extracted cutouts centered on the clusters of interest.

\subsection{Optical and near infrared}\label{sec:optical_data}
\begin{table}[h]
\caption{\footnotesize{Coordinates of the candidate BCGs.}}
\begin{center}
\resizebox{0.5\textwidth}{!} {
\begin{tabular}{c|cc}
\hline
\hline
ID & BCG$_1$ & BCG$_2$ \\
-- & R.A, Dec (deg) & R.A, Dec (deg) \\
\hline
XLSSC~072 & $+33.8500$, $-3.7256^{(a,b)}$ & -- \\ 
XLSSC~100 & $+31.5527$, $-6.1985^{(a)}$   & $+31.5473$ $-6.1920^{(b)}$ \\  
XLSSC~102 & $+31.3196$, $-4.6556^{(a)}$   & $+31.3254$ $-4.6306^{(c)}$\\   
\hline
\end{tabular}
}
\end{center}
{\footnotesize {\bf Notes.} 
$^{(a)}$Paper XXVIII.
$^{(b)}$Paper XV.
$^{(c)}$Identified in Paper XLIV, near the galaxy number density peak; coordinates extracted using the HSC I filter (this work).
}
\label{tab:BCG}
\end{table}

Optical and near-infrared data are used to compare the ICM distribution to that of the galaxies. Here, we followed the procedure presented in Paper XLIV, to which we refer the reader for details. In brief, we used the galaxy photometric catalogs from the Canada-France-Hawai Telescope Legacy Survey \citep[CFHTLS,][]{Gwyn2012} and selected possible member galaxies based on their photometric redshift, magnitude and type (elliptical). We then produced density maps using a Gaussian kernel, leading to a map resolution of 54 arcsec (FWHM of the Gaussian kernel). We then subtracted the contribution from local background and normalized the maps in level of signal-to-background (S/B). To investigate morphology, we used 1000 MC realizations of the galaxy density maps per cluster field, computed by Poissonian realizations of an idealized model fit to the data.

As a complement, we used public data from the Hyper Suprime-Cam Subaru Strategic Program \citep{Aihara2018,Aihara2022} for visual purposes\footnote{\url{https://hsc-release.mtk.nao.ac.jp/doc/}}. In particular, filters R, I, and Z were combined to produce color images of the three target cluster regions.

The locations of the brightest cluster galaxies (BCG) in XXL clusters were identified in \cite{Lavoie2016}, hereafter XXL Paper XV, and \cite{Ricci2018}, hereafter XXL Paper XXVIII, using slightly different criteria. As noted in \cite{Ricci2018PhD} the identified BCGs agree for XLSSC~072 but are different for XLSSC~100, for which we distinguished the two candidate BCGs. The BCG of XLSSC~102 is not reported in Paper XV. However, a second BCG associated with a sub-cluster was discussed in Paper XLIV. We extracted its coordinates using the HSC I filter. The coordinates of the candidate BCGs are listed in Table~\ref{tab:BCG}.

\section{Dynamical state estimates}\label{sec:dynamics}
The ICM pressure profile and the $Y_{\rm SZ} - M$ scaling relation are expected to depend on the cluster dynamical state \citep{Arnaud2010,Yu2015}. It is therefore essential to have information about the dynamical state if one wants to test the standard evolution of these SZ-related observables. In this section, we determined the dynamical state of the clusters according to their morphology and the comparison between the different tracers of the cluster components, including their centers.

\subsection{Morphology}
\begin{figure*}
        \centering
        \includegraphics[width=0.99\textwidth]{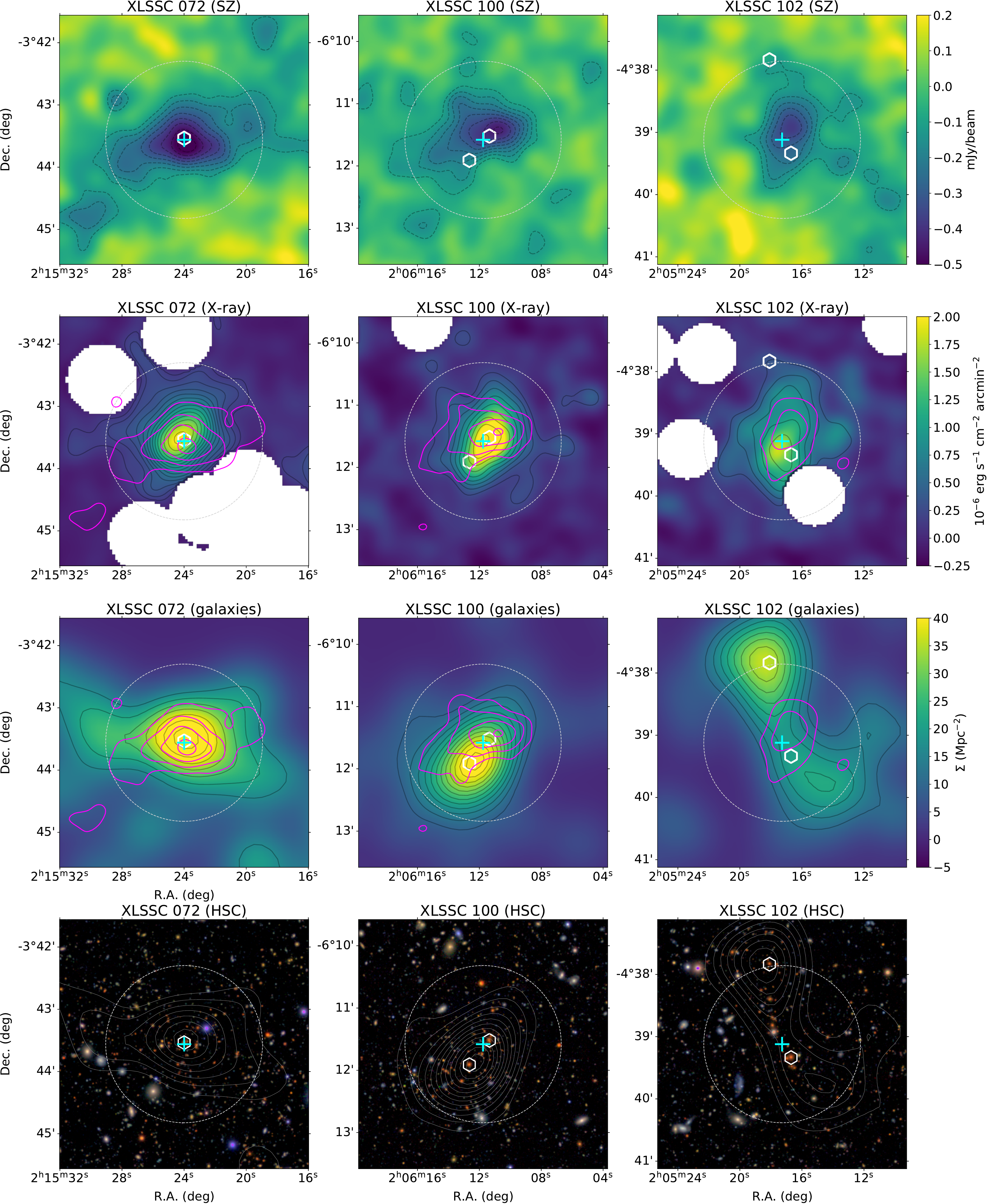}
        \caption{Comparison of SZ, X-ray, and optical data for XLSSC 072 (left), XLSSC 100 (center), and XLSSC 102 (right).
        {\bf First row:} Point-source-subtracted 150 GHz SZ surface brightness images with S/N contours.
        {\bf Second row:} X-ray surface brightness images with S/N contours. Point sources from Paper XXVII have been masked.
        {\bf Third row:} CFHTLS derived galaxy density images, $\Sigma$. 
        {\bf Fourth row:} HSC color images made by combining the R, I, and Z filters.
        In all panels, the cyan cross indicates the reference centers and the BCG positions are indicated as white hexagons.
        The gray dashed circles indicate the characteristic radii $\theta_{500}$ estimated via the $Y_{\rm X}-M$ scaling (see Section~\ref{sec:results}).
    In all panels, the black S/N (or S/B) contours start at $2\sigma$ and are separated by $1\sigma$ each.
        Magenta contours correspond to the SZ S/N, starting at $3\sigma$ and separated by $2\sigma$ each.
    We note that the XLSSC~102 data were already reported in Paper XLIV.
        }
\label{fig:multiwavelenght}
\end{figure*}

The morphology of the clusters is estimated qualitatively via individual SZ, X-ray, and optical data, which we used to trace the ICM thermal pressure, the ICM thermal density, and the galaxy population, respectively. The comparison of these different datasets is also informative given their different sensitivities to the cluster components \citep[e.g.,][]{Komatsu2001,Rasia2013,Nurgaliev2013,Donahue2016,Cialone2018,DeLuca2021}. 

Radio and submillimeter contaminating point sources have been subtracted from SZ images prior to analysis and it should have a negligible impact on the results (see Appendix~\ref{app:point_sources}). X-ray contaminating point sources (essentially AGNs) have been masked conservatively using a 30 arcsec radius aperture according to the catalog presented in \cite{Chiappetti2018}, hereafter XXL Paper XXVII. The X-ray and SZ images have been smoothed to an effective resolution of 27 arcsec (FWHM), while the optical density map kernel is two times larger to ensure a sufficient S/B (54 arcsec FWHM). The R, I, and Z band HSC images were slightly smoothed and combined to produce a color image that visually highlights the cluster member galaxies, given their redshift.

In Figure \ref{fig:multiwavelenght}, we present the multiwavelength view of XLSSC~072, XLSSC~100, and XLSSC~102. This follows the analysis and results presented in Paper XLIV, to which we refer the reader for a more detailed comparison that focuses on XLSSC~102 using a similar strategy. The three clusters are well-detected in all bands. The signal compares well both in terms of amplitude and extension for all sources, in agreement with them having similar masses\footnote{Although we note that the difference in redshift implies a difference in X-ray flux attenuation of up to nearly 30\%.}. Deviation from spherical symmetry is observed in all clusters, which suggests the presence of disturbances in the gas and galaxy distribution. The SZ and X-ray signals overlap well on large scales but may show a significant difference on smaller scales (see Section~\ref{sec:peak_and_centroid_position} for a quantitative comparison of the centroid and peak coordinates) that could indicate local compressions caused by merging events \citep[as in, e.g.,][]{Adam2014}. Both SZ and X-ray signals present relatively flat surface brightness distributions for all clusters, which is consistent with them being dynamically disturbed systems. Accordingly, we do not observe prominent X-ray peaks that would indicate the presence of a dense cool core associated with a relaxed system \citep{Rossetti2010}. As already investigated and reported in Paper XLIV, XLSSC~102 presents a bimodal galaxy number density distribution, while its ICM pressure and density are maximized in between the two peaks. It was interpreted as the result of two merging subclusters. Although the agreement between the galaxy and the gas distribution is better in XLSSC~100, a large offset is observed between the two, with the galaxy density extending more toward the southeast, possibly indicating that the gas is stripped in the direction of a passing subcluster. The ICM and the galaxy density match each other well in XLSSC~072 but they are both elongated in the east-west direction. The presence of multiple BCGs in XLSSC~100 (and possibly XLSSC~102) provides another indication of dynamical activity. In both clusters, one of the BCGs agrees well with the ICM location, while the other is largely offset, as expected for merging clusters with asymmetric mass ratios. Moreover, the elongation of the ICM is roughly aligned with the axis defined by the two BCGs, which agrees with this scenario. In XLSSC~072, the BCG position is well aligned with the ICM and the galaxy distribution center.

All three clusters present morphological signatures of large dynamical activity related to merging events. This is the case in terms of their properties seen in all individual bands for the three clusters and also given the difference observed in the ICM and the galaxy tracers for XLSSC~100 and XLSSC~102. The better agreement between these tracers, in the case of XLSSC~072, could be due to line-of-sight projection effects, in which case the merging event would be mostly oriented along the line-of-sight. According to the qualitative morphological study, we classify XLSSC~100 and XLSSC~102 as dynamically disturbed systems, and XLSSC~072 as likely dynamically disturbed. We note that in Paper XLVIII, XLSSC~072 is classified as disturbed according to the centroid shift estimate, but it would be classified as relaxed according to the BCG-X-ray offset, which is in good agreement with our findings. In Appendix~\ref{app:thermo_profile_diagnosis}, we confirm these conclusions on the cluster dynamical state using the entropy profile, which is another excellent indicator of the ICM thermal state \citep[e.g.,][]{Pratt2010}. 

\subsection{Peak and centroid position}\label{sec:peak_and_centroid_position}

\begin{figure*}
        \centering
        \includegraphics[width=0.32\textwidth]{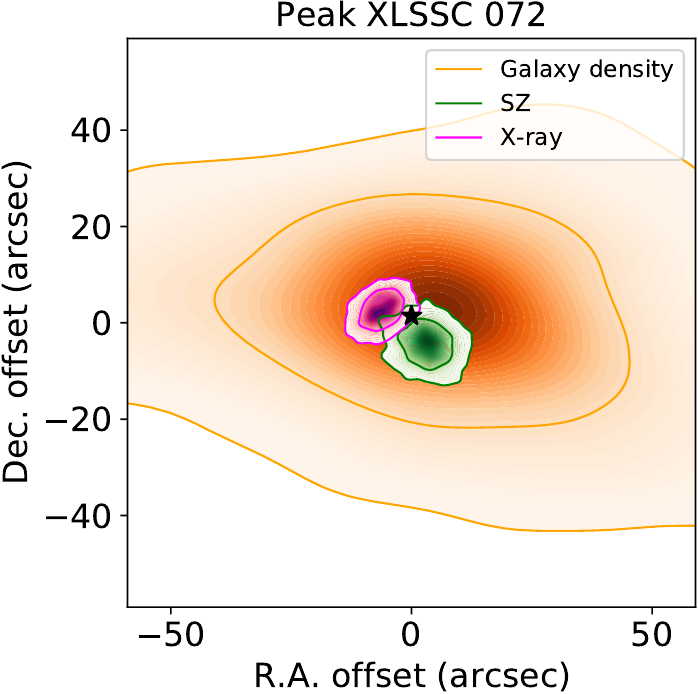}
        \includegraphics[width=0.32\textwidth]{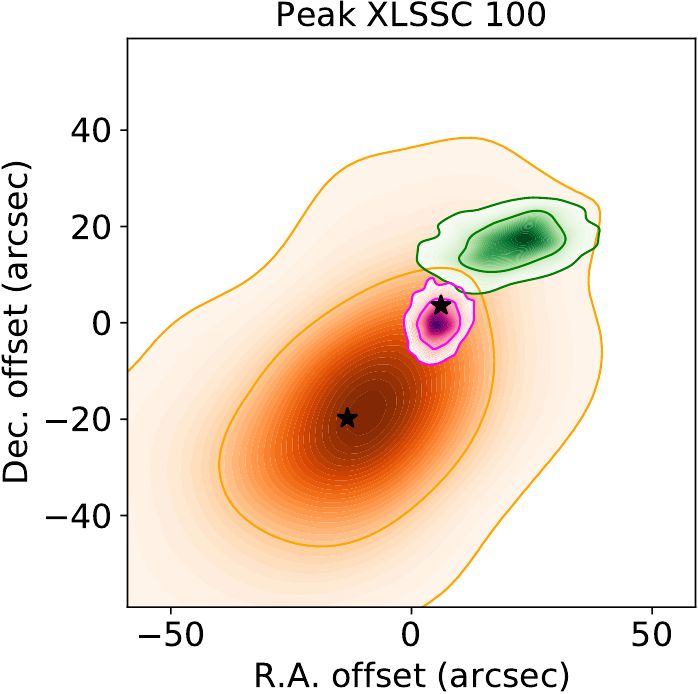}
        \includegraphics[width=0.32\textwidth]{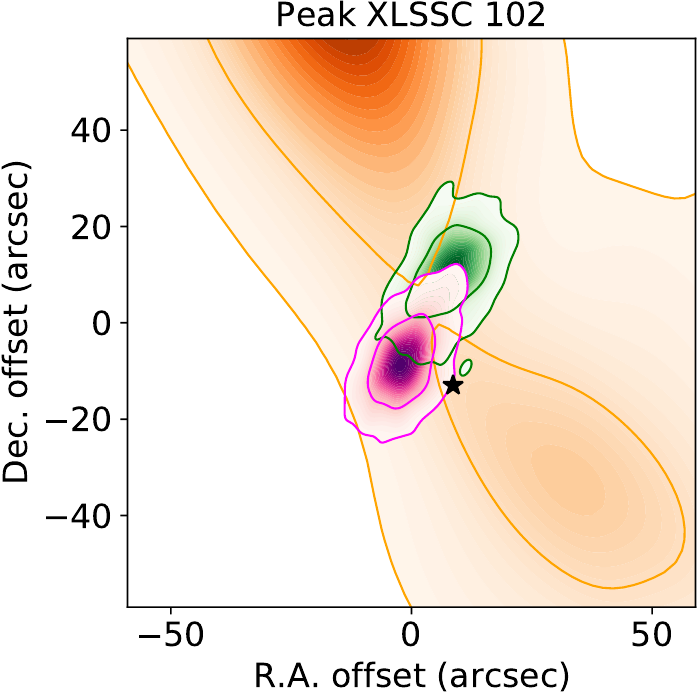}
        \includegraphics[width=0.32\textwidth]{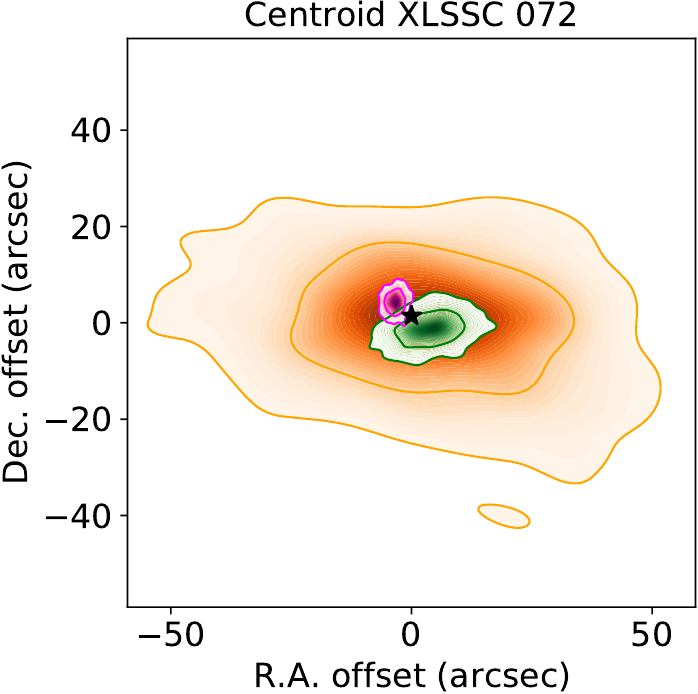}
        \includegraphics[width=0.32\textwidth]{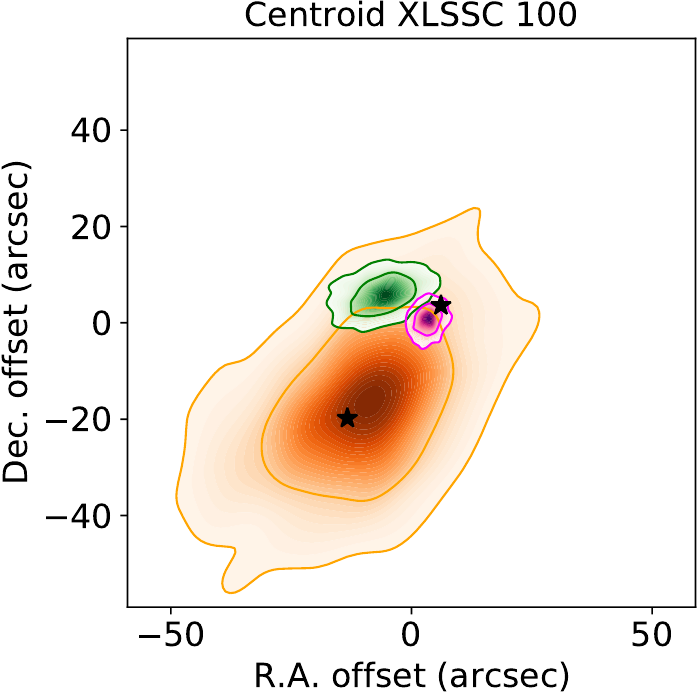}
        \includegraphics[width=0.32\textwidth]{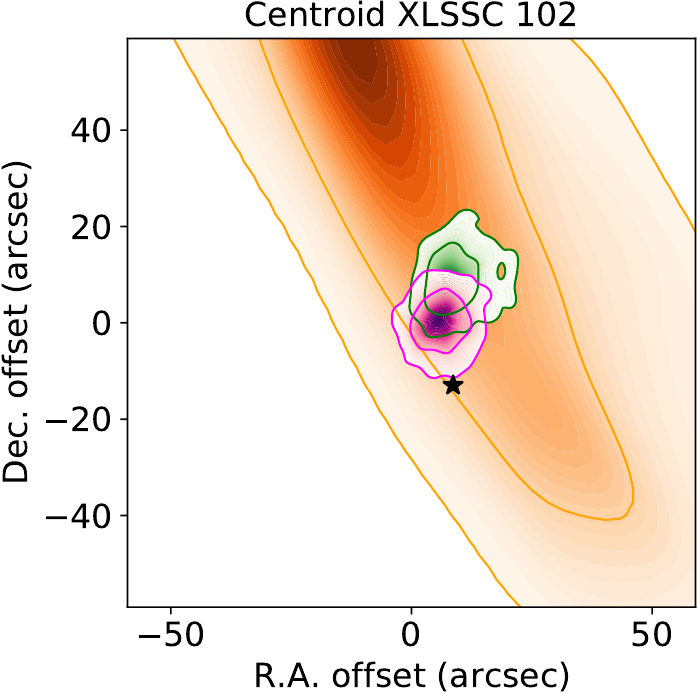}
        \caption{Probability distribution of signal peak (top) and centroid (bottom) location with respect to reference center in three wavelengths, for XLSSC 072, XLSSC 100, and XLSSC 102, from left to right. The BCG coordinates are indicated by the black stars. Contours give the 68\% and 95\% confidence interval. We note that we recover a posterior distribution that is in good agreement with that reported in Paper XLIV for XLSSC~102.}
\label{fig:offset_posterior}
\end{figure*}

The measurement of the offsets between the peaks and centroid of the SZ, X-ray, galaxy density, and the BCGs positions is also an interesting way to address the cluster dynamical state \citep[e.g.,][]{Lin2004,Hudson2010,Rossetti2016,Lopes2018,Zenteno2020,DeLuca2021}. To do so, we reproduced the analysis done for XLSSC~102 in Paper XLIV, but for the full sample. We estimate the peak as the coordinates of the maximum S/N of the signal, taken at the effective resolution of 27 arcsec (FWHM) for the SZ and X-ray data, and 54 arcsec (FWHM) for the galaxy number density. The centroid is obtained by fitting a 2D Gaussian function on the images\footnote{We note that the X-ray centroids roughly coincide with that of the XXL reference coordinates given the detection algorithm (see Table~\ref{tab:sample_summary}).}. Uncertainties are computed by running the same procedure on 1000 MC realizations of each data (except for the BCG coordinates, which have negligible uncertainties). We refer the reader to Paper XLIV for more details on the procedure.

The posterior likelihood distribution in the R.A.-Dec. plane are reported in Figure~\ref{fig:offset_posterior} for the peaks and the centroids, respectively (see also Table~\ref{tab:offset_summary} for numerical results). As expected, better constraints on the centroid are obtained compared to the peak position. Given the uncertainties, the recovered peak and centroid coordinates of XLSSC~072 are consistent for all the data, with the only exception being the tension between the SZ and the X-ray centers, albeit only at a level of about $2 \sigma$. We note that the BCG is located at the intersection of the SZ and X-ray confidence intervals. In the case of XLSSC~100 and XLSSC~102, however, significant differences are observed between the ICM and at least one of the BCGs, both for the peak and the centroid. The other BCG is generally located closer to the X-ray peak and further from the SZ one, in agreement with the scenario in which a merger event induced a local boost of the pressure aside from the remnant denser core of the clusters, next to which the BCG is sitting. In the two clusters, the location of the centroid of the SZ and X-ray agree better than that of the peak, despite the smaller error bars. Again, this supports the fact that a merger event disturbed the cluster cores, while the ICM on large-scale was only weakly affected. In all cases, the uncertainties in the peak and centroid of the galaxy density distributions are too large to draw firm conclusions, although they agree well with the merger scenario by tracking better the BCG that present the largest offset to the ICM.

The offsets between the different cluster components support XLSSC~100 and XLSSC~102 being merging clusters. They also favor -although the evidence is weaker- XLSSC~072 being dynamically perturbed.

\begin{table}[h]
\caption{\footnotesize{Projected physical offsets between the measured peaks and best-fit centroids of different gas and galaxy tracers. The quantity $\Sigma$ refers to the galaxy density. In the case of XLSSC~100 and XLSSC~102, the BCG$_1$ and BCG$_2$ are given in parentheses, respectively. }}
\begin{center}
\resizebox{0.5\textwidth}{!} {
\begin{tabular}{c|cccccc}
\hline
\hline
ID & X-SZ & X-$\Sigma$ & X-BCG & SZ-$\Sigma$ & SZ-BCG & $\Sigma$-BCG  \\
\hline
\multicolumn{7}{c}{Peak (kpc)}\\
\hline
XLSSC~072 &  78   &  69   &  49            & 93   & 46  &  48  \\ 
XLSSC~100 &  101 &  160 &  (215,31)   & 261 & (316, 79)  & (56,183)  \\  
XLSSC~102 &  184 &  667 &   (98, 676) &  540 &  (200, 544) &  (728, 26) \\   
\hline
\multicolumn{7}{c}{Centroid (kpc)}\\
\hline
XLSSC~072 &    74 &  52   &  36               & 39   & 38               &  29  \\ 
XLSSC~100 &    80 &  183 &  (206,32)     & 196  & (210, 91)    & (29, 215)  \\  
XLSSC~102 &    68 &  206 &   (106, 629) & 138  &  (170, 568) &  (308, 441) \\   
\hline
\end{tabular}
}
\end{center}
\label{tab:offset_summary}
\end{table}

\subsection{Search for discontinuities and substructures}
As a complementary investigation of the cluster dynamical state, we searched for substructure and discontinuities in SZ signal using the methodology developed in \cite{Adam2018b} and the Gaussian gradient magnitude and difference of Gaussian filtering. Given the limited S/N of the data and the compactness of the signal at these redshift and mass, we did not find any significant features in the data. This agrees with the results of \cite{Adam2018b}, which state that the signature from merger event is difficult to identify at S/N $\lesssim 10$.

\section{Modeling and analysis procedure}\label{sec:modeling}
This section presents the modeling and analysis methodology developed to extract the pressure profile and the location of our targets on the $Y_{\rm SZ}-M$ relation. After discussing the SZ and X-ray observables, we present the methodology used to extract the density profile, extract the pressure profile, and estimate masses using the hydrostatic equilibrium (HSE) assumption and scaling relations. In this work, we essentially considered HSE masses, despite the fact that our targets present indication for dynamical activity. This choice was motivated by the fact that the cluster pressure profile and the $Y_{\rm SZ}-M$ relation, which we aim to test at high redshift and low mass, were calibrated based on HSE mass measurements (or scaling relations themselves calibrated with HSE masses, e.g., \citealt{Arnaud2010} and \citealt{PlanckXX2014}). Our approach is very similar to these works to ease the comparison. Moreover, given the mass and redshift of our targets, no other reliable individual mass estimates are available. In Appendix~\ref{app:thermo_profile_diagnosis}, we also discuss the reliability of the HSE assumption in light of thermodynamical indicators.

\subsection{Sunyaev-Zel'dovich and X-ray observables}
The SZ effect surface brightness is given by \citep{Birkinshaw1999}
\begin{equation}
\frac{\Delta I_{\nu}}{I_0} = f(\nu) \ y,
\end{equation}
where $I_0$ is the CMB intensity. The characteristic SZ spectrum, $f(\nu)$, does only depend on the frequency in the nonrelativistic approximation, which applies well in the case of our sample. The amplitude of the distortion is given by the Compton parameter, which depends on the line-of-sight integration of the thermal electron pressure, $P_{\rm e}$, as
\begin{equation}
y = \frac{\sigma_{\rm T}}{m_{\rm e} c^2} \int P_{\rm e} d \ell,
\end{equation}
where $\sigma_{\rm T}$ is the Thomson cross-section and $m_{\rm e} c^2$ the electron rest mass. Given the NIKA2 beam and bandpass at 150 GHz, a Compton parameter $y = 10^{-4}$ corresponds to a surface brightness of $\Delta I_{\nu} = -1.19 \pm 0.09$ mJy/beam \citep{Ruppin2018}.

The X-ray surface brightness is expressed as \citep{Sarazin1986}
\begin{equation}
S_{\rm X} = \frac{1}{4 \pi \left(1+z\right)^4} \int n_e^2 \Lambda(T,Z) d \ell,
\end{equation}
where $n_e$ is the thermal gas electron number density. The cooling function, $\Lambda$, depends weakly on the temperature, $T$, and on the metallicity, $Z$.

\subsection{Extraction of the thermal electron density profile}\label{sec:Extraction_thermal_density_profile}
The cluster electron density profiles are extracted as in Paper XLIV. In brief, we used the \textsc{pyproffit} package \citep{Eckert2020}\footnote{\href{https://pyproffit.readthedocs.io}{https://pyproffit.readthedocs.io}}, which is the Python implementation of the \textsc{proffit} software \citep{Eckert2011}. The X-ray surface brightness was extracted in radial bins of 5 arcsec width by accumulating photon counts within each annulus and correcting the vignetting by dividing by the local exposure map. As in Section~\ref{sec:dynamics}, point sources from the XXL catalog were masked by excluding circles of 30 arcsec radius, corresponding to an encircled energy fraction of $\sim90\%$. The multi-scale decomposition developed in \cite{Eckert2016} (hereafter XXL Paper XIII) was used to deproject the thermal electron number density profile assuming spherical symmetry. A single-temperature APEC model \citep{apec} absorbed by the Galactic $N_H$ was used to convert from X-ray count rate to emission measure, with temperature fixed to the ones from Paper III listed in Table~\ref{tab:sample_summary}. The model was convolved with the \textit{XMM-Newton} point spread function and fitted to the data using a Poisson likelihood in \textsc{PyMC} with the No U-Turn Sampler \citep{pymc}. In the end, we obtained the best-fit electron number density profile together with 1000 realizations that we used to compute uncertainties. We note the presence of small differences compared to the profile presented in Paper XLIV. They are due to better modeling of the point spread function, accounting for the exact location of the cluster in the field of view and the combination of the multiple XXL pointings. The profiles are reported in Appendix~\ref{app:density_profiles}.

\subsection{Extraction of the thermal pressure profile}\label{sec:Extraction_thermal_pressure_profile}
We modeled the ICM thermal pressure via the \textsc{MINOT} software \citep{Adam2020} using several different approaches described hereafter. \textsc{MINOT} allows us to easily produce SZ maps, given a pressure profile, projected on the same grid as the data. The maps are convolved with the NIKA2 beam and the transfer function that describes the filtering induced by the data reduction procedure \citep{Adam2015}. The surface brightness profiles are extracted in bins of 5 arcsec in width and up to a distance of 3 arcmin from the cluster center. As a reference, the XXL detection center is used, corresponding roughly to the X-ray centroid. We discuss this choice in Section~\ref{sec:results}. Given a model of the pressure profile, the parameters are fitted to the data using a Markov chain Monte Carlo (MCMC) technique. The parameter space is sampled with the \textsc{emcee} package \citep{Foreman2013}. A Gaussian likelihood function was used to compare the model and the data. We account for the full covariance matrix, computed using MC realizations of the noise (see \citealt{Adam2016} for the procedure). In addition to the NIKA2 data, we also impose a Gaussian prior on the total SZ flux (see Equation~\ref{eq:Ysz}, integrated up to $5 R_{500}$). To do this, we use the Planck measurement obtained by fitting a Gaussian function with a 10 arcmin FWHM (i.e., the Planck beam size) on the Compton parameter map \citep{PlanckXXII2016} at the location of the unresolved targets (see Table~\ref{tab:Yplanck}). As for the density profiles, we propagate the uncertainties on the pressure profile using 1000 pressure profile models randomly taken from the MCMC chains.

We consider the two following ways of modeling the pressure profile (see also Section~\ref{sec:Direct_HSE_Mass_estimates} for the direct modeling of the mass profile, which is also an alternative way of modeling the pressure profile indirectly).

The first is the Generalized Navarro-Frenk-White model. As a baseline, the pressure is described according to the generalized Navarro-Frenk-White (gNFW) model \citep{Nagai2007}, expressed as
\begin{equation}
P_e(r) = \frac{P_0}{\left(\frac{r}{r_p}\right)^c \left(1 + \left(\frac{r}{r_p}\right)^a\right)^{\frac{b-c}{a}}}.
\label{eq:gNFW}
\end{equation}
The parameter $P_0$ is a normalization, $r_{p}$ is a characteristic radius, and $c$, $a$, and $b$ describe the slope of the profile from the core to the outskirts. The fit parameters include all the pressure profile parameters from Equation~\ref{eq:gNFW} plus the map zero level as a nuisance parameter. We use flat priors on the normalization and scale radius ($P_0>0$ and $5 R_{500}>r_p>0$, with $R_{500}$ measured from the $Y_X-M$ relation; see Section~\ref{sec:Scaling_Mass_estimates}) and Gaussian priors on the slope parameters based on \cite{PlanckV2013}: $\mu_{a,b,c} = [1.33, 4.13, 0.31]$ and $\sigma_{a,b,c} = [1.00, 3.10, 0.23]$, corresponding to 75\% of the mean value (i.e., 3 times larger than the values used in Paper XLIV, to ensure more freedom in the profile shape).

We also consider the modeling of the pressure by fixing the normalization of the profile at given radii \citep[see, e.g.,][for a similar method]{Ruppin2017}. This is the binned model. The full profile is then interpolated in logarithmic space before line-of-sight integration and projection, as implemented in \textsc{MINOT}. We define the values of the radii as five bins logarithmically spaced from 50 kpc to 1 Mpc, $r_i \equiv [50, 106, 224, 473, 1000]$ kpc. This allows us to sample the profiles where NIKA2 is sufficiently sensitive and obtain reliable constraints in each bin. The five pressure model parameters are given by $P_i \equiv P(r_i)$, to which the zero level of the map is added as a nuisance parameter. The pressure parameters are restricted to verify $P_i > 0$ in all bins $i$.\\

\begin{table}[h]
\caption{\footnotesize{Planck prior on the total flux.}}
\begin{center}
\resizebox{0.25\textwidth}{!} {
\begin{tabular}{c|c}
\hline
\hline
ID & $D_A^2 Y_{{\rm SZ}, {\rm tot}}$ (kpc$^2$)  \\
\hline
XLSSC~072 &  $22 \pm 61$  \\  
XLSSC~100 &  $64 \pm 62$  \\  
XLSSC~102 &  $39 \pm 55$  \\  
\hline
\end{tabular}
}
\end{center}
\label{tab:Yplanck}
\end{table}

\subsection{Direct hydrostatic equilibrium mass estimates}\label{sec:Direct_HSE_Mass_estimates}
Assuming that the ICM is in hydrostatic equilibrium and spherically symmetric, the total mass enclosed within the radius $r$ is given by
\begin{equation}
M_{\rm HSE}(r) = -\frac{r^2}{G \mu_{\rm gas} m_{\rm p} n_{\rm e}(r)} \frac{d P_e(r)}{dr},
\label{eq:MHSE}
\end{equation}
where $\mu_{\rm gas}=0.61$ is the gas mean molecular weight, $m_{\rm p}$ is the proton mass, and $G$ the Newton constant. We extract the hydrostatic equilibrium mass profile using the two following methods.

First, the HSE mass profile given in Equation~\ref{eq:MHSE} is obtained by combining the density and the pressure profiles measured independently from the X-ray and SZ data, as discussed in Sections~\ref{sec:Extraction_thermal_density_profile} and~\ref{sec:Extraction_thermal_pressure_profile}, respectively. Given the critical density of the Universe at the cluster's redshift, we derive the overdensity contrast by integrating the mass profile, which we use to obtain $R_{500}$ and thus compute $M_{{\rm HSE},500}$. Uncertainties are propagated from the pressure and the density profiles by combining 1000 model realizations randomly taken from the MCMC chains.

Alternatively, we directly model the total mass density profile as the sum of the gas density, which we know from X-ray data (see Equation~\ref{eq:gas_mass_profile}), and a Navarro-Frenk-White \citep[NFW,][]{Navarro1996} model to describe the other components (essentially the dark matter). 
We note that in practice, we have checked that modeling the total mass with a single NFW model does not significantly affect our results since the gas is absorbed in the NFW component. The NFW density model is written as
\begin{equation}
\rho_{\rm NFW}(r) = \frac{\rho_0}{\left(\frac{r}{r_s}\right) \left(1 + \frac{r}{r_s}\right)^2}.
\label{eq:NFW}
\end{equation}
In the case where the NFW model accounts for the total mass (gas included), the characteristic radius can be simply written as $r_s = R_{500}/c_{500}$ and
\begin{equation}
\rho_0 = \frac{500 \rho_c c_{500}^3}{3 \left({\rm log}\left(1+c_{500}\right) - \frac{c_{500}}{1+c_{500}}\right)},
\end{equation}
with $\rho_c$ being the critical density of the Universe at the cluster's redshift and $c_{500}$ the concentration. The enclosed hydrostatic mass is given by
\begin{equation}
M_{\rm HSE}(r) = 4 \pi \rho_0 r_s^3  \left({\rm log}\left(\frac{r_s+r}{r_s}\right)-\frac{r}{r_s+r}\right) + M_{\rm gas}(r).
\label{eq:NFW}
\end{equation}
Taking advantage of the \textsc{MINOT} code implementation, we combine this model with the density profile and Equation~\ref{eq:MHSE} to compute the pressure profile model,
\begin{equation}
P_e(r) = P_e(r_0) + \int_r^{r_0} \frac{G \mu_{\rm gas} m_{\rm p} n_e(r^{\prime}) M_{\rm HSE}(r^{\prime})}{{r^{\prime}}^2}dr^{\prime},
\label{eq:NFW}
\end{equation}
with $r_0$ being a radius taken sufficiently far, at which point the pressure is negligible\footnote{We note that $P_e(r_0)$ is fully degenerate with the zero level of the map so that it can be ignored in the fit.}. The pressure profile model is then compared to the NIKA2 data as in Section~\ref{sec:Extraction_thermal_pressure_profile}. However, in this case, the density profile is randomly sampled from the 1000 MC realization available at each step of the MCMC to account for the associated uncertainty. The fit parameters that we use are the mass $M_{\rm HSE, 500}$ and the concentration $c_{500}$, which are related to $\rho_0$ and $r_s$. While the main goal of this method is to directly describe the mass profile with a physically motivated model, this also provides an alternative pressure profile model that complements the methods described in Section~\ref{sec:Extraction_thermal_pressure_profile}. We refer the reader to \cite{Eckert2022} and \cite{Munoz2023}, for example, for a detailed description of this approach.

\subsection{Mass estimates from scaling relations and fitting the universal pressure profile}\label{sec:Scaling_Mass_estimates}
In addition to direct mass measurements based on the HSE assumption, it is useful to compute and compare masses estimated using the global cluster properties that are usually easier to obtain. We thus also use the UPP normalization as a mass proxy.

\subsubsection{Mass estimation from the universal pressure profile}
The UPP \citep[e.g.,][which we follow here, or any other calibration of the profile]{Arnaud2010} depends exclusively on the cluster mass and redshift. In this scenario, Equation~\ref{eq:gNFW} can be expressed as
\begin{equation}
P_e(r) = P_{500} f_M \frac{P_0}{\left(c_{500} \frac{r}{R_{500}}\right)^c \left(1 + \left(c_{500} \frac{r}{R_{500}}\right)^a\right)^{\frac{b-c}{a}}},
\label{eq:UPP}
\end{equation}
with $P_{500} \equiv P_{500}(M_{500})$ being the self-similar normalization \citep{Nagai2007} and $f_M = \left(\frac{M_{500}}{3 \times 10^{14} {\rm M}_{\odot}}\right)^{0.12}$ a small mass dependence correction. Following \cite{Arnaud2010}, the parameters of the profile are set to $\left(P_0, c_{500}, a, b, c\right) = \left(8.403, 1.177, 1.0510, 5.4905, 0.3081\right)$. We use Equation~\ref{eq:UPP} as in Section~\ref{sec:Extraction_thermal_pressure_profile} to fit the NIKA2 data with the mass $M_{500}$ and the map zero level as the only free parameters. This is similar to the methodology used by \cite{Hilton2018,Hilton2021} to extract ACT masses. Given the fact that the clusters appear as dynamically active, we also reproduce this work by using the mean profile of morphologically disturbed clusters from \cite{Arnaud2010}. We note that the UPP was calibrated using nearby clusters and assumes standard evolution, which is what we aim to test in the present work, as we discuss in Section~\ref{sec:results}. Although they are not directly obtained from the HSE assumption, the masses used in the UPP calibration were obtained from the $Y_{\rm X}-M$ relation, itself calibrated using the direct HSE masses of relaxed clusters \citep{Arnaud2007}.

\subsubsection{Mass estimates from the $Y_{\rm SZ}-M$ relation}
The SZ flux $Y_{{\rm SZ}, 500}$ is tightly correlated with the mass. The best-fit scaling relation as calibrated in \cite{PlanckXX2014} is given by
\begin{equation}
E(z)^{-2/3} \left(\frac{D_A^2 Y_{{\rm SZ}, 500}}{10^{-4} {\rm Mpc}^2}\right) = 10^{-0.19} \times \left(\frac{M_{{\rm HSE},500}}{6\times 10^{14} {\rm M}_{\odot}}\right)^{1.79}.
\label{eq:YSZ-M_scaling}
\end{equation}
It was obtained using masses derived using the $Y_{\rm X}-M$ relation, itself calibrated using HSE masses computed from X-ray observations. This relation is used to estimate the mass according to our SZ flux measurement. To do so, we compute the spherically integrated SZ flux given the pressure profile as
\begin{equation}
Y_{{\rm SZ}}(R) = \frac{\sigma_{\rm T}}{m_e c^2} \int_0^{R} 4 \pi r^2 P_e(r) dr.
\label{eq:Ysz}
\end{equation}
Equation (\ref{eq:Ysz}) is integrated up to $R_{500}$ to obtain $Y_{{\rm SZ},500}$, and thus depends on $M_{500}$. Therefore, we perform the measurement by iterating about the scaling relation (convergence is obtained within less than 1\% after a few iterations). The full probability distribution in the $Y_{\rm SZ}-M$ plane is obtained by repeating the measurement with 1000 pressure profiles randomly taken from the MCMC chains. By default, we use the pressure profile obtained from the gNFW fit to the data. We note that although it is possible to use Equation (\ref{eq:YSZ-M_scaling}) to estimate the cluster's mass, this relation, as calibrated using nearby clusters, is precisely what we aim to test in the present work. We discuss it in Section~\ref{sec:results}.

\subsubsection{Mass estimates from the $Y_{\rm X}-M$ relation}
The X-ray analog of the SZ flux, $Y_{{\rm X},500} \equiv k_{\rm B} T_{\rm X} M_{{\rm gas},500}$ \citep{Kravtsov2006}, is an excellent mass proxy \citep{Arnaud2007,Vikhlinin2009b}. Here, we use the best-fit scaling relation from \cite{Arnaud2010} in order to obtain a high-quality mass estimate based on direct X-ray-only measurement:
\begin{equation}
E(z)^{-2/3} \left(\frac{Y_{{\rm X}, 500}}{2 \times 10^{14} {\rm M}_{\odot} {\rm keV}}\right) = 10^{0.376} \times \left(\frac{M_{{\rm HSE},500}}{6\times 10^{14} {\rm M}_{\odot}}\right)^{1.78}.
\label{eq:Yx-M_scaling}
\end{equation}
The masses used to calibrate this relation were obtained applying the HSE on relaxed clusters observed in X-ray. We estimate $Y_{{\rm X},500}$ using the measured X-ray temperatures listed in Table~\ref{tab:sample_summary} (Paper III). The gas mass profile is computed from the gas density profile as 
\begin{equation}
M_{\rm gas}(R) = \int_0^R 4 \pi r^2 \mu_{\rm e} m_{\rm p} n_{\rm e}(r) dr,
\label{eq:Mgas}
\end{equation}
with $\mu_{\rm e} = 1.16$ being the electron mean molecular weight. As for $Y_{\rm SZ, 500}$, the integration is performed up to $R_{500}$ to obtain $M_{{\rm gas},500}$. Since the estimate of $Y_{{\rm X}, 500}$ depends on $R_{500}$, and thus $M_{500}$, we perform the measurement by iterating about the scaling relation. The full probability distribution, and thus the uncertainty on the mass, is obtained by repeating the measurement with 1000 density profiles taken from the MC realizations and simultaneously sampling the temperature within its uncertainty. Again, as discussed in Section~\ref{sec:results}, we note that Equation (\ref{eq:Yx-M_scaling}) was calibrated using nearby clusters and assumes standard evolution, which is what we aim at testing in the present work.

\section{Results and discussions}\label{sec:results}
In this section, we present the results of the analysis. After discussing the mass measurements, we focus on the pressure profile and the $Y_{\rm SZ}-M$ scaling relation.

\subsection{Masses}\label{sec:masses}
\begin{figure*}
        \centering
        \includegraphics[width=0.99\textwidth]{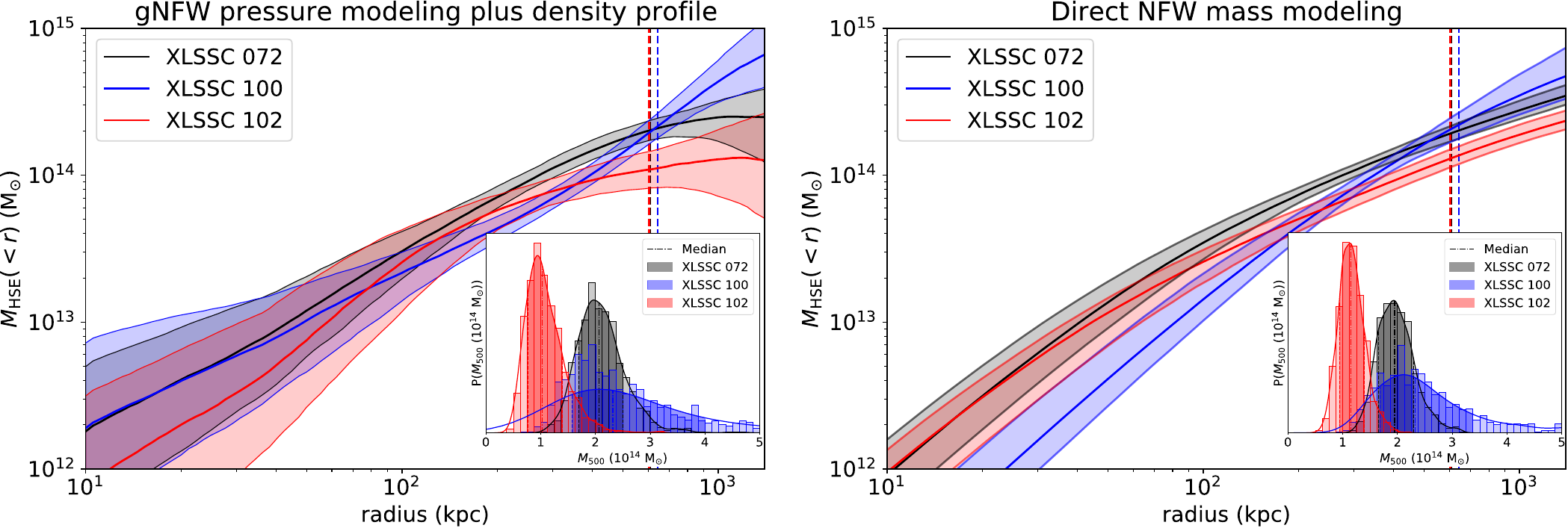}
        \caption{HSE mass profiles of XLSSC~072, XLSSC~100 and XLSSC~102.
        {\bf Left:} Mass profile obtained using the gNFW pressure model together with the electron density profile.
        {\bf Right}: Mass profile obtained by direct NFW mass modeling together with the electron density profile.
        For reference, the vertical dashed lines represent $R_{500}$ estimated using the $Y_{\rm X}-M$ relation.
        The corresponding probability density functions for $M_{500}$ are shown as insets in the figures. 
        The shaded region gives a 68\% confidence interval.
        }
\label{fig:mass_profile_constraints}
\end{figure*}

\subsubsection{Direct HSE mass profiles}
Figure~\ref{fig:mass_profile_constraints} presents the HSE mass profiles obtained either from combining the density profiles with the gNFW pressure model, or directly modeling the mass with an NFW profile. The NFW model leads to the smallest uncertainties due to fewer parameters involved, but the two methods show excellent agreement over the full radial range, highlighting the robustness of the measurement. The quality of the recovered profiles is remarkable given the low masses and the high redshifts of these clusters.

The three clusters present comparable mass profiles. They flatten at $r \gtrsim R_{500}$ for XLSSC~072 and XLSSC~102, but it keeps rising for XLSSC~100. This feature is due to XLSSC~100 having a flatter outer pressure profile and a slightly steeper outer density profile than the other two clusters. It implies significantly larger uncertainties on the recovered value of $M_{500}$ for XLSSC~100 than XLSSC~072 and XLSSC~102. The numerical results on the mass are reported in Table~\ref{tab:Y-M}, where we also give the corresponding SZ flux. We refer to Appendix~\ref{app:thermo_profile_diagnosis} for further investigation of the reliability of the mass profile using thermodynamics diagnosis.

\subsubsection{Comparison between direct measurements, estimates from scaling laws, and the literature}
\begin{figure*}
        \centering
        \includegraphics[width=0.99\textwidth]{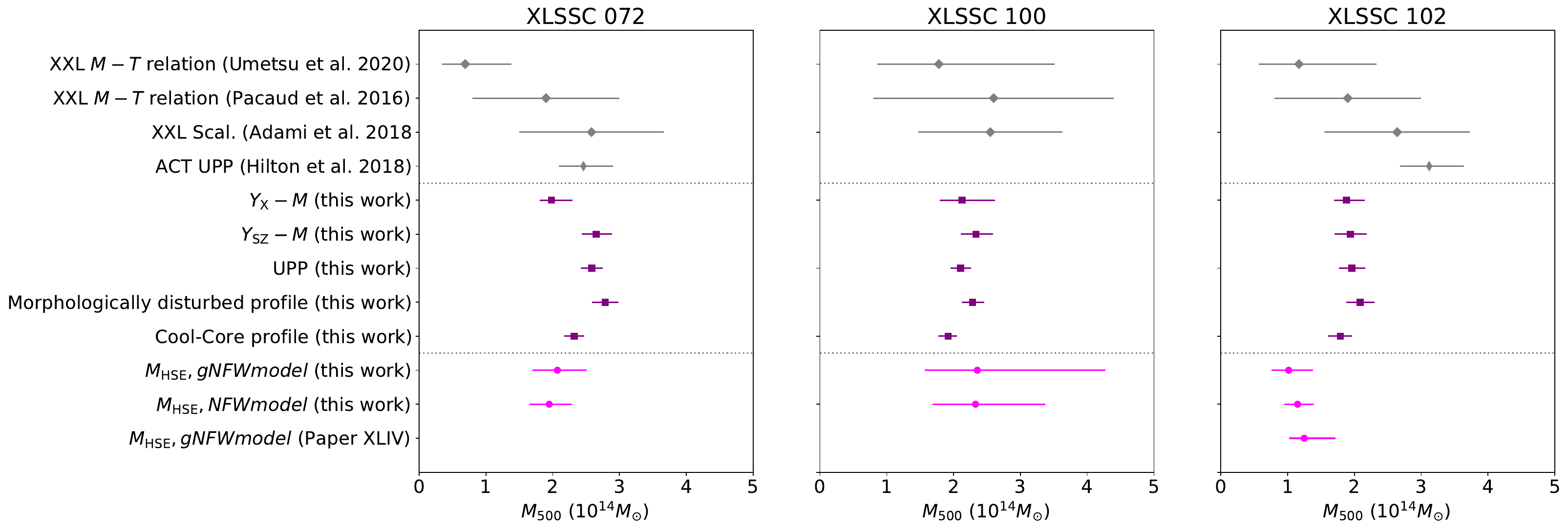}
        \caption{Comparison between different mass measurements reported in Table~\ref{tab:sample_summary} and Table~\ref{tab:Y-M}. The first block (gray points) corresponds to survey measurements (from XXL and ACT), the second block to masses derived using low-redshift, higher mass calibration proxies in the present work (purple points), and the last block to direct HSE measurements from the present work (magenta points). We also report the value from Paper XLIV for XLSSC~102 when assuming a similar method and center. The small difference is mainly due to the updated X-ray density profile used in the present work.
        }
\label{fig:mass_comparison}
\end{figure*}

\begin{table}[h]
\caption{\footnotesize{SZ fluxes and masses. The median value and the 68\% confidence interval are reported. All the masses are assimilated to hydrostatic masses (see Section~\ref{sec:modeling} for the different methodologies).}}
\begin{center}
\resizebox{0.5\textwidth}{!} {
\begin{tabular}{c|cc}
\hline
\hline
 & $D_A^2 Y_{{\rm SZ}, 500}$ (kpc$^2$) & $M_{500}$ ($10^{14}$M$_{\odot}$)  \\
\hline
 & \multicolumn{2}{c}{Direct HSE masses (gNFW pressure model + density)}\\
\hline
XLSSC~072 &  $20.1_{-3.3}^{+4.2}$ & $2.07_{-0.37}^{+0.44}$   \\  
XLSSC~100 &  $17.3_{-4.5}^{+8.0}$ & $2.35_{-0.79}^{+1.92}$   \\  
XLSSC~102 &  $9.9_{-2.2}^{+3.1}$ & $1.02_{-0.26}^{+0.37}$   \\  
\hline
 & \multicolumn{2}{c}{Direct HSE masses (NFW mass modeling + density)}\\
\hline
XLSSC~072 &  $20.3_{-3.3}^{+3.2}$ & $1.95_{-0.30}^{+0.34}$   \\  
XLSSC~100 &  $16.6_{-3.5}^{+5.0}$ & $2.33_{-0.64}^{+1.05}$   \\  
XLSSC~102 &  $11.2_{-2.3}^{+2.3}$ & $1.15_{-0.20}^{+0.25}$   \\  
\hline
 & \multicolumn{2}{c}{Universal pressure profile fit}\\
\hline
XLSSC~072  & $19.8^{+2.4}_{-2.2}$ & $2.58_{-0.17}^{+0.17}$   \\   
XLSSC~100 &  $13.2^{+1.8}_{-1.8}$ & $2.11_{-0.15}^{+0.15}$  \\   
XLSSC~102 &  $12.1^{+2.1}_{-2.2}$ & $1.96_{-0.20}^{+0.20}$   \\   
\hline
 & \multicolumn{2}{c}{Morphologically disturbed pressure profile fit}\\
\hline
XLSSC~072  & $23.2^{+2.8}_{-3.0}$ & $2.78^{+0.20}_{-0.19}$   \\   
XLSSC~100 &  $15.7^{+1.9}_{-2.1}$ & $2.28^{+0.17}_{-0.16}$  \\   
XLSSC~102 &  $13.7^{+2.5}_{-2.6}$ & $2.08^{+0.22}_{-0.21}$   \\   
\hline
 & \multicolumn{2}{c}{Cool-core pressure profile fit}\\
\hline
XLSSC~072  & $16.6^{+1.8}_{-2.0}$ & $2.32^{+0.15}_{-0.15}$   \\   
XLSSC~100 &  $11.4^{+1.3}_{-1.7}$ & $1.92^{+0.13}_{-0.14}$  \\   
XLSSC~102 &  $10.2^{+1.8}_{-2.0}$ & $1.79^{+0.18}_{-0.18}$   \\   
\hline
 & \multicolumn{2}{c}{$Y_{\rm SZ}-M$ scaling relation}\\
\hline
XLSSC~072 &  $21.8_{-3.1}^{+3.7}$ & $2.65_{-0.21}^{+0.24}$   \\  
XLSSC~100 &  $16.8_{-2.8}^{+3.4}$ & $2.34_{-0.23}^{+0.25}$   \\  
XLSSC~102 &  $12.3_{-2.6}^{+2.9}$ & $1.94_{-0.24}^{+0.24}$   \\  
\hline
 & \multicolumn{2}{c}{$Y_{\rm X}-M$ scaling relation}\\
\hline
XLSSC~072 &  $19.9_{-2.4}^{+2.6}$ & $1.98_{-0.17}^{+0.31}$   \\ 
XLSSC~100 &  $16.2_{-2.4}^{+2.7}$ & $2.13_{-0.33}^{+0.49}$   \\ 
XLSSC~102 &  $12.3_{-2.2}^{+2.3}$ & $1.88_{-0.19}^{+0.28}$   \\ 
\hline
\end{tabular}
}
\end{center}
\label{tab:Y-M}
\end{table}

The masses derived with the methods presented in Section~\ref{sec:Direct_HSE_Mass_estimates} are listed in Table~\ref{tab:Y-M}. Only the direct HSE masses are independent of any calibration at low redshift. We also report the SZ flux enclosed within $R_{500}$. In Figure~\ref{fig:mass_comparison}, we compare the masses derived in the present work to those obtained in the literature (Table~\ref{tab:sample_summary}).

The masses obtained from XXL scaling relations reflect the large uncertainty in the mass proxies on which they rely and the precision of the scaling relation. This is also the case for our $Y_{\rm X}-M$ and $Y_{\rm SZ}-M$ masses, although they rely on more precise mass proxies given the data in hand and on scaling relations that are expected to be very tightly related to the mass. We note that these relations are calibrated using X-ray data \citep{Arnaud2010}, but we do not correct for any mean HSE bias here. The masses derived through the UPP normalization should match the $Y_{\rm SZ}-M$ ones perfectly in the case of a perfect UPP profile. The HSE masses that we derive are the only direct measurements. However, they are affected by systematics in the modeling and by the hydrostatic mass bias. In Appendix~\ref{app:thermo_profile_diagnosis}, we discuss possible biases in the recovered masses in light of thermodynamics diagnosis.

In Figure~\ref{fig:mass_comparison}, we observe a very good general agreement between the different mass measurements despite the very different methodologies and assumptions involved. Only XLSSC~102 presents a $\gtrsim 2 \sigma$ tension between the direct HSE masses and the masses obtained from the $Y_{\rm X}-M$ and $Y_{\rm SZ}-M$ scaling relations and the UPP normalization fit. It could be due to the ongoing merger activity that affects the HSE assumption. In Appendix~\ref{app:thermo_profile_diagnosis}, we show that the HSE masses are likely to be biased low, by up to a factor of 2 for XLSSC~102, which would reconcile the different estimates. Focusing on the purple points, good agreement is obtained between $Y_{\rm X}-M$ and $Y_{\rm SZ}-M$ masses. This indicates that the $Y_{\rm X}-Y_{\rm SZ}$ relation followed by our targets is in excellent agreement with the one measured in \cite{Arnaud2010}. In the case of XLSSC~072, the observed difference vanishes when using the temperature reported in the detailed analysis of Paper XLVIII instead of the one from Paper III, when computing $Y_{\rm X}$ (see also Appendix~\ref{app:thermo_profile_diagnosis}). The UPP-based masses that we derive agree very well with the $Y_{\rm SZ}-M$ masses, which indicates that the shape of the pressure profiles does not strongly deviate from that of the UPP. However, an $\sim 2 \sigma$ tension between our UPP-based mass and that obtained from ACT data \citep{Hilton2018} is observed for XLSSC~102 despite the similar methodology employed. When changing the UPP to morphologically disturbed or cool-core models, the changes in the mass are about $1\sigma$.

In the following, we use these masses to compare our NIKA2 measurements with the pressure profile and $Y_{\rm SZ}-M$ expected from standard evolution. Given the precision in the masses and the underlying assumptions that they involve, we use the $Y_{\rm X}-M$ and the direct HSE (NFW-based) masses for reference.

\subsection{Pressure profile}\label{sec:results_pressure_profile}
\begin{figure*}
        \centering
        \includegraphics[width=0.99\textwidth]{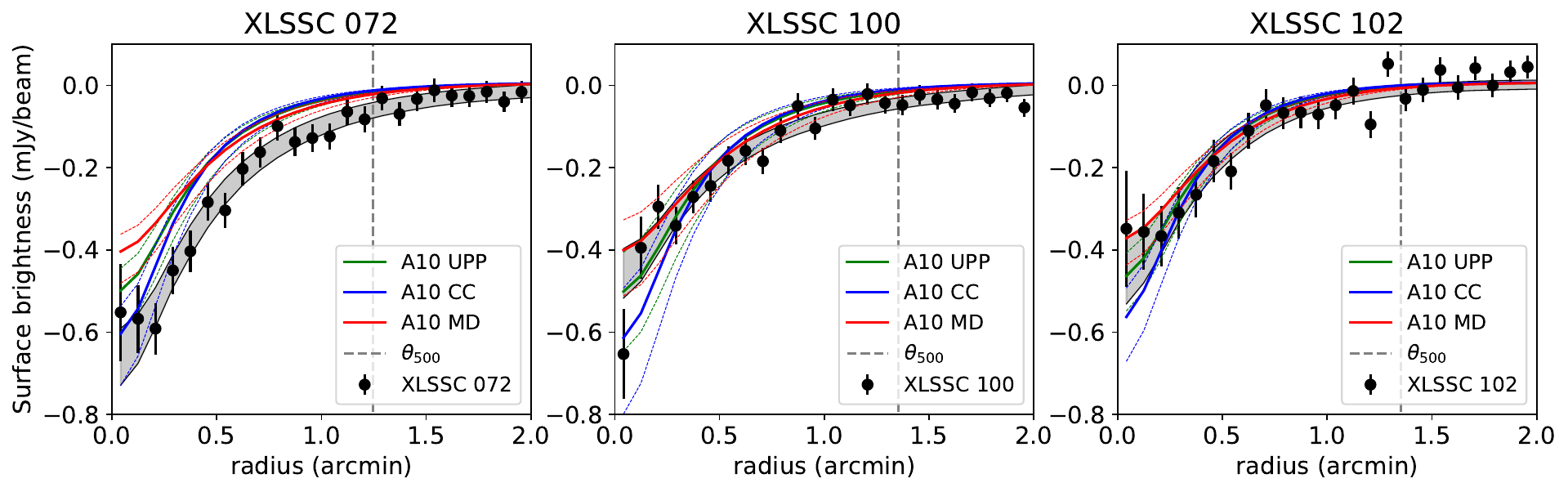}
        \includegraphics[width=0.99\textwidth]{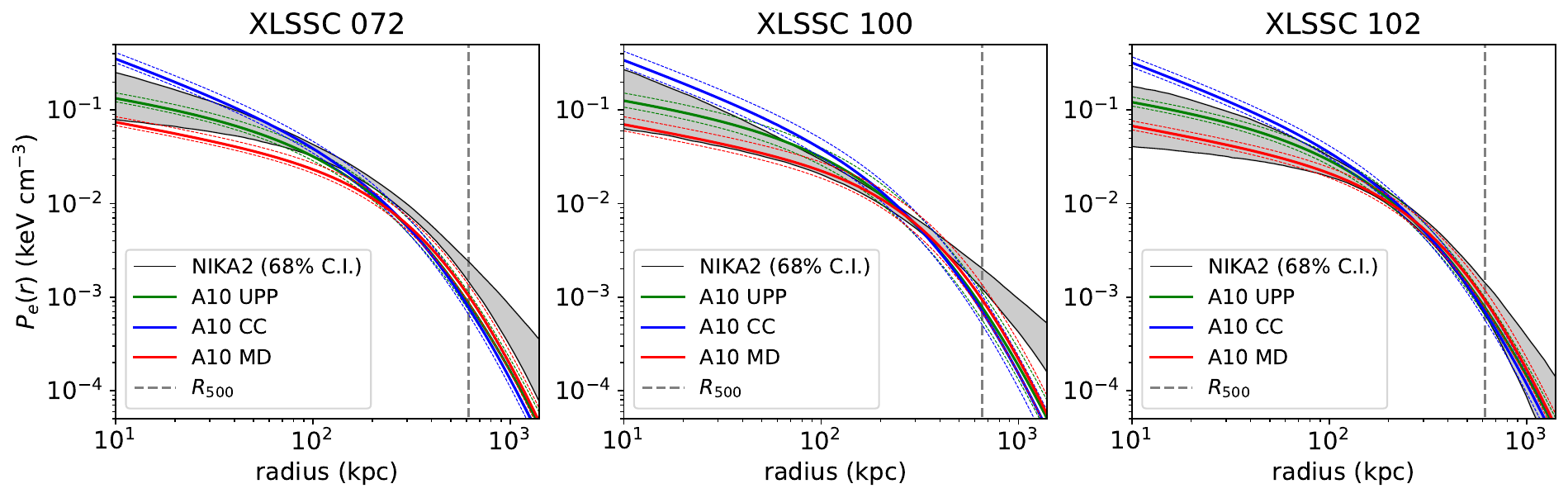}
        \includegraphics[width=0.99\textwidth]{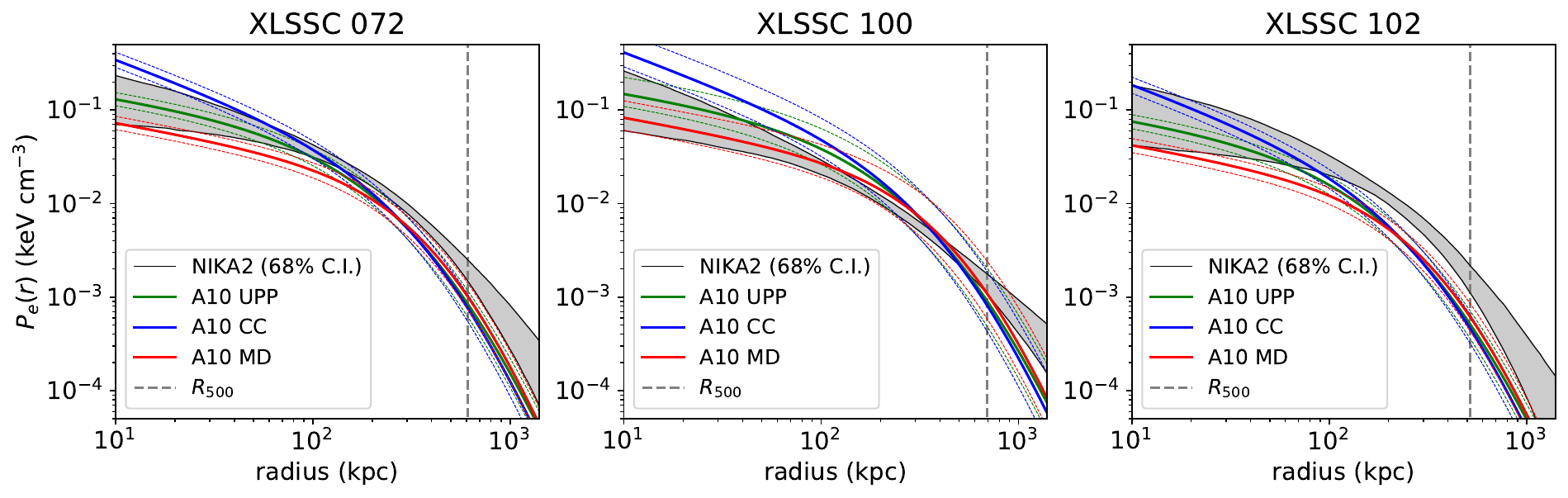}
        \caption{NIKA2 constraints on the thermal pressure profile.
    {\bf Top:} gNFW constraints on the SZ surface brightness.
    {\bf Middle}: gNFW constraints on the thermal electron pressure profile.
        The gray band provides the $1\sigma$ constraint on the model. The green, blue, and red lines give the expected model according to the UPP, the cool-core (CC) pressure profile, and the morphologically disturbed (MD) pressure profile according to \cite{Arnaud2010} given the $Y_{\rm X}-M$ masses. The corresponding dashed lines give the same model assuming higher or lower masses by $1 \sigma$. 
    {\bf Bottom}: Same as the middle row, but computing the expected models given the masses derived from the direct NFW fit to the pressure profile plus the density profile.
        }
\label{fig:pressure_constraints}
\end{figure*}

\begin{figure*}
        \centering
        \includegraphics[width=0.99\textwidth]{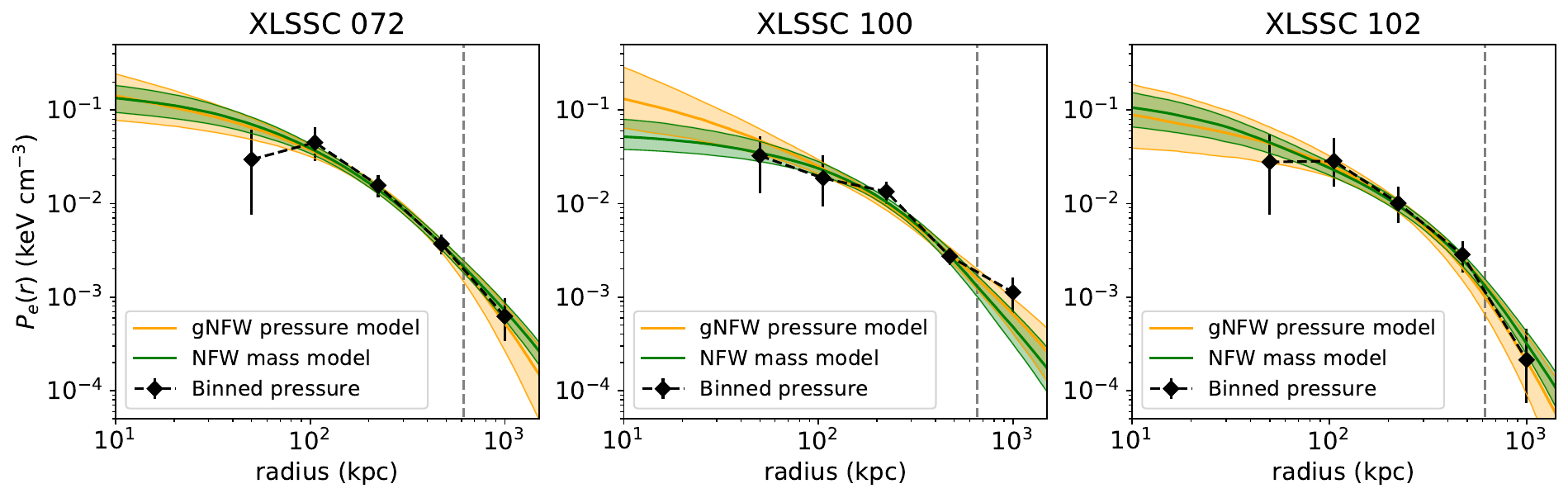}
        \caption{Comparison of pressure profile as measured with different methods: gNFW modeling of the pressure, binned pressure profile, and NFW modeling of the mass with the joint use of the density profile. The vertical dashed lines give the location of $R_{500}$ as in Figure~\ref{fig:mass_comparison}.}
\label{fig:pressure_constraints_methods}
\end{figure*}

The SZ surface brightness profiles of XLSSC~072, XLSSC~100, and XLSSC~102 and their corresponding pressure profiles are shown in Figure~\ref{fig:pressure_constraints} for the case of the gNFW model. The SZ decrement is detected up to about $R_{500}$ for XLSSC~072 and XLSSC~100. The S/N is slightly lower for XLSSC~102. The best-fit models describing the data and their uncertainties are obtained as discussed in Section~\ref{sec:modeling}. We report the 68\% interval allowed by the data as a gray band. While we use the full covariance matrix in the analysis, the uncertainties only provide the diagonal of the covariance matrix. We refer the reader to Paper XLIV for more details about the computation of the covariance matrix. The residuals between the best-fit model and the data are provided in Appendix~\ref{app:residual_SZ} at the map level.

The comparison with the models from \cite{Arnaud2010}, namely the UPP, the mean morphologically disturbed profile, and the mean cool-core profile, is performed using $Y_{\rm X}-M$ and direct NFW masses. The surface brightness models have been convolved with the instrument response function. The derived pressure profiles reflect the same behavior but are deconvolved from the instrument response and projection effects. All the measured profiles agree best with the morphologically disturbed model in terms of shape. They are significantly different than the averaged cool-core pressure profile, but they still agree with the UPP within error bars. The choice of the mass is highly relevant to the comparison in terms of amplitude. For instance, the models describing XLSSC~102 reach $2\sigma$ lower when using direct HSE mass measurements, while the match is excellent with the $Y_{\rm X}-M$ mass. Similarly, a much better match would be obtained for XLSSC~072 by using the temperature from Paper XLVIII instead of the one from Paper III to compute $Y_{\rm X}$. As already mentioned regarding the mass profile, XLSSC~100 presents a pressure profile outer slope that is more shallow than expected from the models at the $\sim 2 \sigma$ level.

Assuming standard evolution, the shape of the pressure profile is in excellent agreement with the dynamical state analysis of Section~\ref{sec:dynamics}. It suggests that XLSSC~072, XLSSC~100, and XLSSC~102 are increasingly disturbed systems, with even XLSSC~072 showing evidence of disturbance. Alternatively, given the prior knowledge of the cluster dynamical states inferred in Section~\ref{sec:dynamics}, the data are in good agreement with the pressure profile calibrated on low-redshift clusters \citep{Arnaud2010} and scaled to low mass and high redshift using standard evolution. The agreement would be even better when accounting for the intrinsic scatter in the expected profile.

In Figure~\ref{fig:pressure_constraints_methods}, we compare the pressure profiles recovered using the three methods described in Section~\ref{sec:modeling}. Despite the very different methodologies, all profiles show excellent agreement within uncertainties at all radii. In the case of these systems, the NIKA2 data are most sensitive to the pressure profile in the range from about 100 kpc to 600 kpc. It is remarkable that reliable constraints on the pressure profile can be obtained nearly up to $2 R_{500}$ for such high-redshift and low-mass clusters.

\subsection{The $Y_{\rm SZ}-M$ relation}
\begin{figure*}
        \centering
        \includegraphics[width=0.99\textwidth]{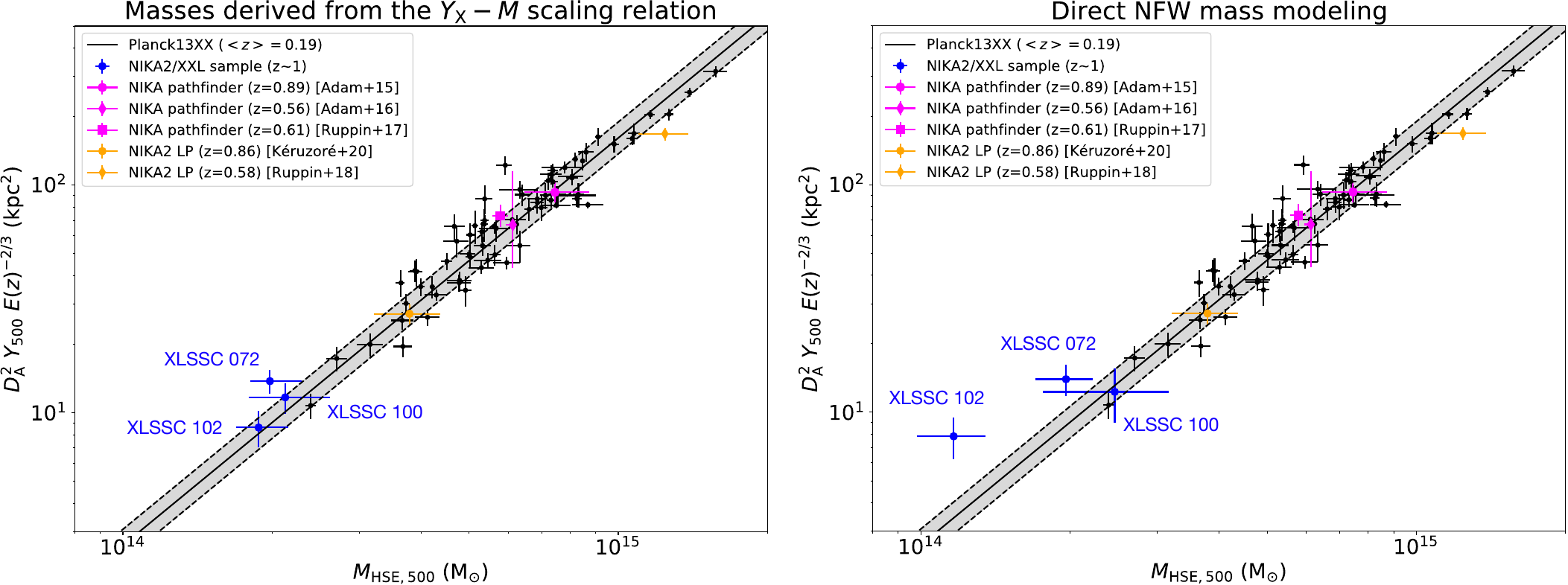}
        \caption{Scaling relation between the SZ flux and the cluster mass.
        The blue points are those obtained in this work, at redshift $z \sim 1$ and $M_{500} \sim 2 \times 10^{14}$ M$_{\odot}$.
        The black points correspond to the Planck calibration sample from \cite{PlanckXX2014}, at a mean redshift of $z = 0.19$. The gray band provides the best-fit relation and the intrinsic scatter.
        Other individual measurements obtained with NIKA and NIKA2 are reported as indicated in the legend.
    We note that the systematic uncertainty associated with the center definition and pressure substructure reported in Paper XLIV is comparable to the size of the error bars for XLSSC~102.
        {\bf Left:} Masses obtained from $Y_{\rm X}-M$ relation.
        {\bf Right:} Masses obtained from HSE assumption (NFW mass modeling).
        }
\label{fig:ym}
\end{figure*}

Figure~\ref{fig:ym} compares the $Y_{\rm SZ}-M$ relation followed by XLSSC~072, XLSSC~100, and XLSSC~102, to the Planck calibration sample used to derive cosmological constraints \citep{PlanckXX2014}. The \cite{PlanckXX2014} relation was obtained from a sample of $z < 0.45$ clusters, with a mean redshift $\left<z\right>=0.19$. The masses were derived using the $Y_{\rm X} - M$ relation from \cite{Arnaud2010}, calibrated using an X-ray sample of 20 local clusters. The masses used for the calibration are thus HSE masses, as given in Equation~\ref{eq:Yx-M_scaling}. The measured intrinsic scatter (7\%) is reported as the gray band on the figure, together with the best-fit relation (Equation~\ref{eq:YSZ-M_scaling}). For comparison, we also show the location of other NIKA and NIKA2 observed clusters \citep[$0.5<z<0.9$,][]{Adam2015,Adam2016,Ruppin2017,Ruppin2018,Keruzore2020} on the relation, but we stress that the flux and masses were not derived in a homogeneous way for those. The masses (and thus SZ fluxes, see Section~\ref{sec:modeling}) used for the XXL sample are either direct HSE masses obtained from the combination of the NIKA2 and XMM-Newton data, or those obtained using the $Y_{\rm X} - M$ relation from \cite{Arnaud2010}. Therefore, in the latter case, we implicitly tested the $Y_{\rm SZ} - Y_{\rm X}$ relation at high redshift and low mass.

As we can observe, our sample sits at the low-mass end of the \cite{PlanckXX2014} calibration sample, but our clusters are located at redshift $z \sim 1$ instead of $z \sim 0.2$. Nonetheless, thanks to the quality of the data, we were able to obtain comparable uncertainties on the flux, but uncertainties on the mass remain larger by a factor of two or more. Despite the different regime that we probed and the fact that these clusters are significantly disturbed \citep[implying a likely higher intrinsic scatter,][]{Yu2015}, the XXL clusters follow the scaling relation remarkably well when using the $Y_{\rm X}-M$ relation to obtain the mass. This is also the case for other NIKA and NIKA2 clusters with published SZ fluxes and masses, at higher masses and lower redshifts. When using direct HSE mass measurement, only XLSSC~102 deviates from the relation by about $2 \sigma$. However, as investigated in detail in Paper XLIV and Appendix~\ref{app:thermo_profile_diagnosis}, this may be related to systematic uncertainties in the mass measurement due to the very complex morphology and dynamical state of this cluster or a very large hydrostatic mass bias.

Either way, we do not observe any significant deviation from the $Y_{\rm SZ} - M$ scaling relation in the three high-redshift, low-mass XXL clusters. Moreover, our results implicitly show that the three clusters follow the $Y_{\rm SZ} - Y_{\rm X}$ relation remarkably
well. While the size of our sample does not allow us to infer statistical conclusions on the relation, this provides a first indication of these relations being stable down to $M_{500} \sim 2 \times 10^{14}$ M$_{\odot}$ and $z \sim 1$.

\subsection{Discussion}\label{sec:discussions}
The results presented in the paper may be affected by the analysis choices, which we discuss here. For instance, we used the XXL detection center as the reference for extracting the profiles and derived quantities. While not much freedom is available for XLSSC~072 given the agreement between the different cluster components on the center, this is not the case for XLSSC~100 and XLSSC~102. Paper XLIV explored the systematic uncertainty associated with this choice for the most perturbed cluster of our sample, XLSSC~102. Hence, this provides us with an upper limit on this uncertainty for the sample, which is in fact modest for the global quantities ($M_{\rm HSE, 500}$, $Y_{\rm SZ, 500}$) -on the order of $\frac{1}{2} \sigma$- but it can be as high as about $1 \sigma$ for the profiles in the central region. Similarly, the morphological analysis showed that the clusters are not spherically symmetric. By performing the analysis in different sectors for XLSSC~102, Paper XLIV estimated the corresponding dispersion to be $\gtrsim 1 \sigma$ on the profiles, but slightly smaller on global quantities. 

Our analysis also relies on the modeling of the pressure profile or the HSE mass profile. Nevertheless, we tested that different methodologies relying on very different assumptions led to consistent results. Therefore, the systematic uncertainty associated with the modeling is expected to be much smaller than statistical uncertainties. The direct HSE masses that we derived rely on spherical symmetry and the HSE assumption and are more likely to be affected by systematic effects (cluster geometry, clumping, etc.; see Appendix~\ref{app:thermo_profile_diagnosis}) than indirect methods, but they are the only direct measurement that can be used to test the $Y_{\rm SZ}-M$ relation. On the other hand, the most robust and precise masses are likely to be the ones derived from the $Y_{\rm X}-M$ relation, the $Y_{\rm SZ}-M$ relation, or the UPP normalization fit, but they rely on the calibration of these relations at lower redshifts and higher masses, which is what we aimed to test here. Moreover, these methods do not generally propagate the intrinsic scatter associated with the scaling relation or pressure profile.

The main conclusion of our work is that the pressure profile and the SZ mass proxy are in line with standard extrapolation down to $z \sim 1$ and $M_{500} \sim 2 \times 10^{14}$ M$_{\odot}$. However, this is based on a sample of only three clusters. Stronger conclusions would require increasing the size of the sample, but this might require a significant amount of observing time for such masses and redshifts, which is not straightforward to obtain. Despite the limitation of the present work, the results indicate that the physics that drives cluster formation is already in place in the regime that we explored.

\section{Summary and conclusions}\label{sec:Summary_and_conclusions}
The SZ structure of the ICM gives us precious information about the thermal state of galaxy clusters and the astrophysical processes at play during their formation. This is reflected in the cluster thermal pressure profile and the scaling relation maintained by the SZ flux and their mass. Detailed investigations of these properties have been done at low redshift, and the effort is being put in at high redshifts for massive clusters. However, at high redshifts and low masses, where the largest deviations from self-similarity are expected, the investigation of the SZ structure with resolved data has remained nearly unexplored to date.

In this paper, we present the analysis of three XXL-selected clusters at $z \sim 1$ and $M_{500} \sim 2 \times 10^{14}$ M$_{\sun}$ observed with the NIKA2 camera via their SZ signal, at a resolution of about 18 arcsec. We investigated the dynamical state of the sources using SZ, X-ray, and optical data. We extracted their pressure profile and compared them to expectations from standard evolution. Complementary X-ray data were used to extract the gas density profile, which we combined with the pressure to measure the hydrostatic masses of the systems. We also estimated the masses using the UPP normalization fit to the data,  $Y_{\rm SZ} - M$, and the $Y_{\rm X} - M$ scaling relations.

The main conclusions of this work are listed here.
\begin{itemize}
\item The three clusters, XLSSC~072, XLSSC~100, and XLSSC~102, at $z \sim 1$ and $M_{500} \sim 2 \times 10^{14}$ $M_{\odot}$ are well detected with NIKA2 in about hours hours per source. The signal is extended and the peak S/N reaches $-9.7$, $-9.2$, and $-6.9$, respectively. These are among the first resolved SZ data available down to such low masses and high redshifts.
\item All three clusters present evidence for ongoing merging activity. This is shown by their disturbed morphologies that present deviation from a compact, spherically symmetric distribution. In the case of XLSSC~100 and XLSSC~102, this is further confirmed by the offset between the peak and centroid of the SZ, the X-ray, the galaxy density, and the BCGs. Assuming standard evolution, this is also confirmed by the flatness of their pressure profiles.
\item The pressure profile is well constrained up to $2 R_{500}$, which is a remarkable achievement given the low masses and high redshifts of the clusters.
\item The pressure profile of the three clusters agrees with that of local dynamically disturbed systems, once rescaled according to standard evolution in mass and redshift. In the case of XLSSC~072, the data are in better agreement with expectations from dynamically disturbed systems, but they also agree with the UPP.
\item Despite their perturbed ICM, their low masses, and high redshifts, we do not find any significant deviation in the $Y_{\rm SZ}-M$ scaling relation followed by our targets.
\item The comparison of the pressure profile and the $Y_{\rm SZ} - M$ scaling relation to that of local samples is limited by uncertainties in the mass. This highlights the difficulty of obtaining accurate and robust mass estimates in this new regime.
\end{itemize}
Galaxy cluster formation is primarily driven by gravity, on top of which feedback processes help regulate cluster evolution. This includes shock heating, turbulent cascade of energy injected from large-scale structures accretion and mergers, and supernova and AGN feedback. These processes are expected to shape the radial thermodynamical profiles and scaling relations followed by galaxy clusters. The lack of significant nonstandard evolution in the pressure profile and the $Y_{\rm SZ}-M$ relation of clusters when extrapolating those expected from lower redshift and more massive objects suggests that the dominant mechanisms that drive clusters' observational properties are already in place around $z \sim 1$, down to $M_{500} \sim 10^{14}$M$_{\odot}$.


\begin{acknowledgements}
We are thankful to the anonymous referee for useful comments, which helped improve the quality of the paper.

MP acknowledges long-term support from the Centre National d'Etudes Spatiales.
We thank L. Chiappetti for his careful reading  of the paper.

This work was supported by the Programme National Cosmology et Galaxies (PNCG) of CNRS/INSU with INP and IN2P3, co-funded by CEA and CNES.

This work is based on observations carried out under project number 179-17, 094-18, 208-18, 093-19, 218-19, and 076-20 with the NIKA2 camera at the IRAM 30 m Telescope. IRAM is supported by INSU/CNRS (France), MPG (Germany) and IGN (Spain).

We would like to thank the IRAM staff for their support during the campaigns. 
The NIKA dilution cryostat has been designed and built at the Institut N\'eel. In particular, we acknowledge the crucial contribution of the Cryogenics Group, and  in particular Gregory Garde, Henri Rodenas, Jean Paul Leggeri, Philippe Camus. 
This work has been partially funded by the Foundation Nanoscience Grenoble, the LabEx FOCUS ANR-11-LABX-0013 and the ANR under the contracts 'MKIDS' and 'NIKA'. 
This work has benefited from the support of the European Research Council Advanced Grants ORISTARS and M2C under the European Union's Seventh Framework Programme (Grant Agreement nos. 291294 and 340519).

Based on observations obtained with XMM-Newton, an ESA science mission with instruments and contributions directly funded by ESA Member States and NASA

XXL is an international project based around an XMM Very Large Programme surveying two $25$ deg$^2$ extragalactic fields at a depth of $\sim 6\cdot 10^{-15} {\rm erg}\cdot {\rm cm}^{-2}{\rm s}^{-1}$ in the [0.5--2] keV band for point-like sources. The XXL website is \url{http://irfu.cea.fr/xxl}. Multi-band information and spectroscopic follow-up of the X-ray sources are obtained through a number of survey programmes, summarized at \url{http://xxlmultiwave.pbworks.com}. 

This study is based on observations obtained with MegaPrime/MegaCam, a joint project of CFHT and CEA/IRFU, at the Canada-France-Hawaii Telescope (CFHT) which is operated by the National Research Council (NRC) of Canada, the Institut National des Science de l'Univers of the Centre National de la Recherche Scientifique (CNRS) of France, and the University of Hawaii. This work is based in part on data products produced at Terapix available at the Canadian Astronomy Data Centre as part of the Canada-France-Hawaii Telescope Legacy Survey, a collaborative project of NRC and CNRS. 

This paper present data collected at the Subaru Telescope and retrieved from the HSC data archive system, which is operated by Subaru Telescope and Astronomy Data Center at National Astronomical Observatory of Japan. Data analysis was in part carried out with the cooperation of Center for Computational Astrophysics, National Astronomical Observatory of Japan.

This research made use of Astropy, a community-developed core Python package for Astronomy \citep{Astropy2013}, in addition to NumPy \citep{VanDerWalt2011}, SciPy \citep{Jones2001}, and Ipython \citep{Perez2007}. Figures were generated using Matplotlib \citep{Hunter2007}. 
\end{acknowledgements}

\bibliographystyle{aa}
\bibliography{biblio_XXL}

\begin{appendix}

\section{Calibration of XLSSC 072 data}\label{app:xlssc072_calib}
\begin{figure}
        \centering
        \includegraphics[width=0.45\textwidth]{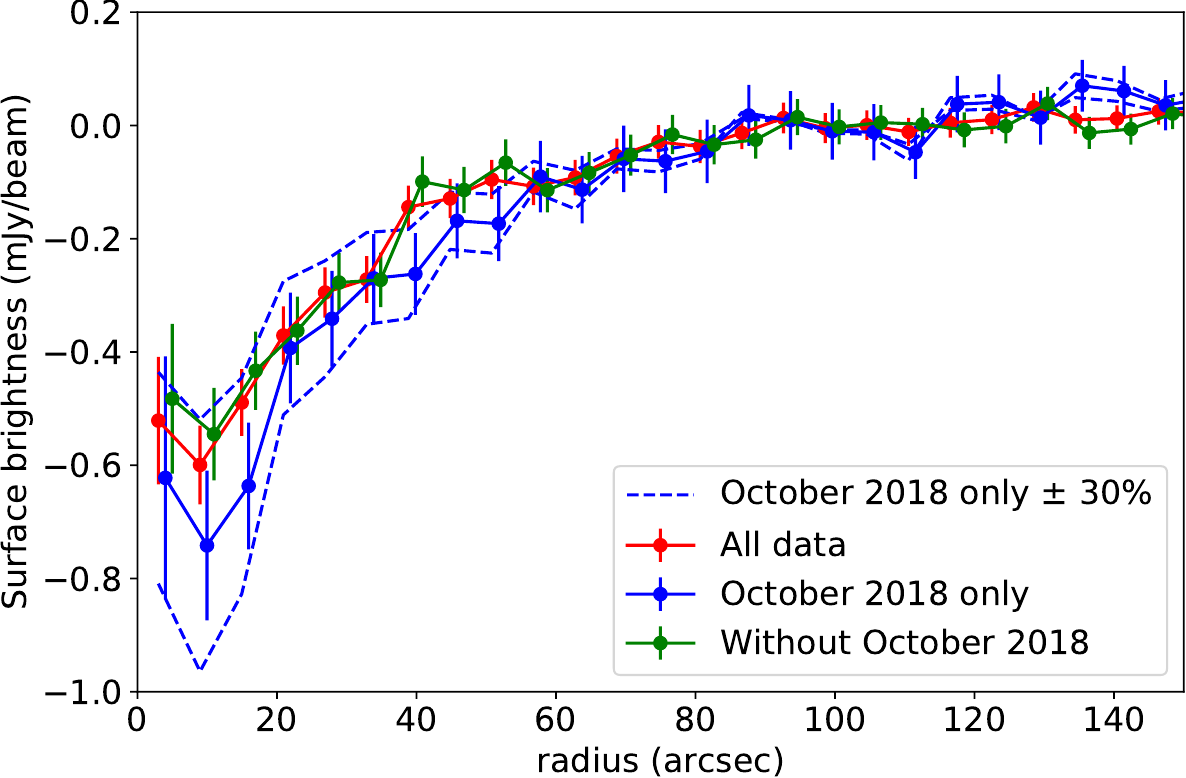}
        \caption{Surface brightness profile of XLSSC~072, after point source subtraction, for the different datasets and their combination.}
\label{fig:xlssc072_profile_valide}
\end{figure}

About a quarter of the data obtained toward XLSSC~072 (October 2018) were calibrated using a bright radio source in the cluster field, instead of the standard method, because of a failure in the calibration system. In Figure~\ref{fig:xlssc072_profile_valide}, we present the cluster surface brightness profile for the different dataset. We check that the SZ profiles are in agreement within the statistical uncertainties and the 30\% calibration uncertainty expected for the October 2018 data.

\section{Radio and submillimeter point source identification and modeling}\label{app:point_sources}

\begin{figure*}
        \centering
        \includegraphics[width=0.99\textwidth]{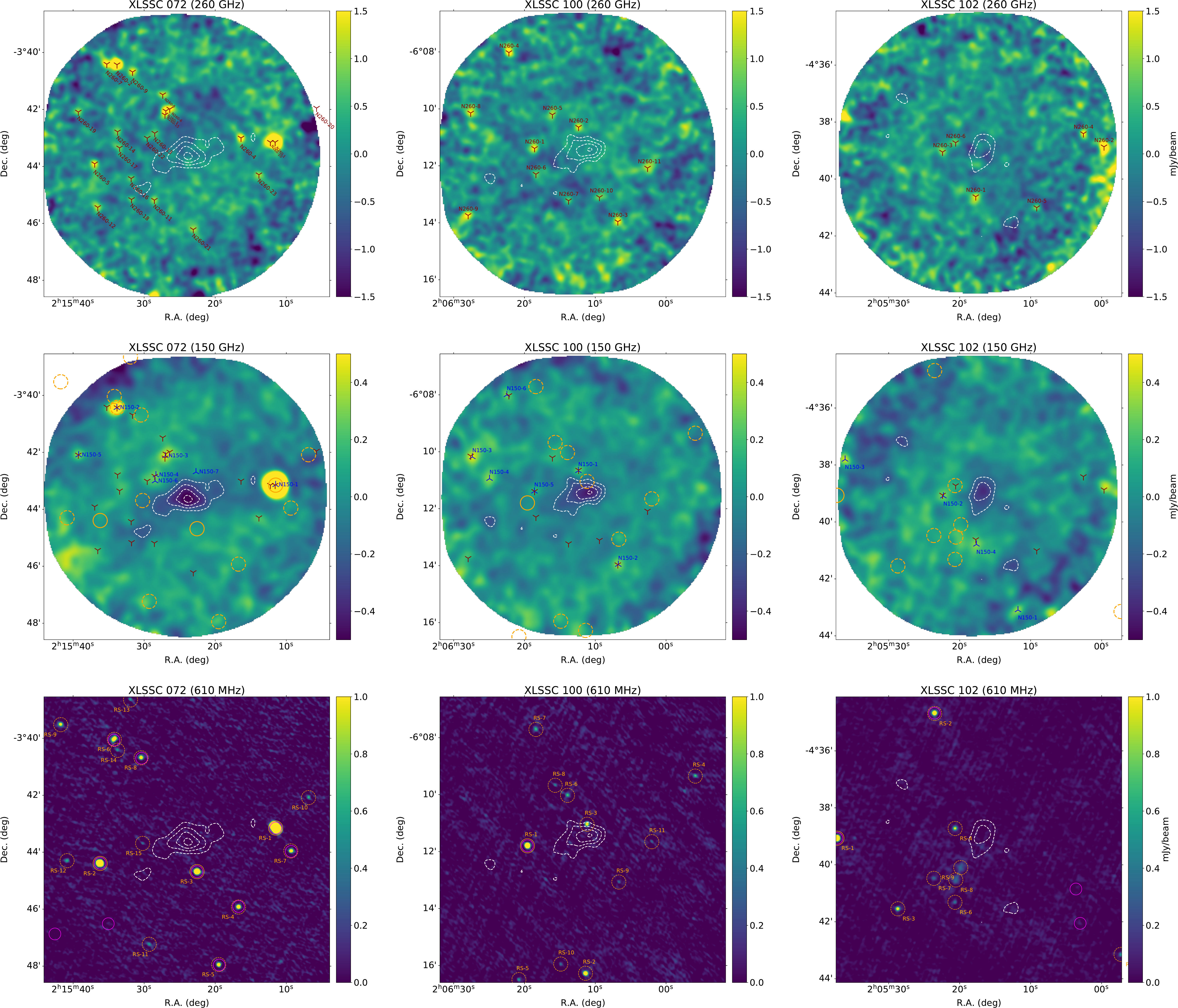}
        \caption{Submillimeter and radio sources' identification in the fields of XLSSC~072 (left), XLSSC~100 (center), and XLSSC~102 (right).
        {\bf Top:} NIKA2 260 GHz images with detected sources indicated as red crosses.
        {\bf Middle:} NIKA2 150 GHz images with detected sources indicated as blue crosses.
        {\bf Bottom:} GMRT images at 610 MHz, with detected sources indicated as orange circles. Sources with NVSS counterparts are indicated as solid lines, and dashed otherwise. FIRST sources are indicated as magenta circles.
        The 150 GHz S/N contours at -9,-7,-5, and -3 $\sigma$ are reported in all maps.
        All identified radio and submillimeter sources are reported in the 150 GHz maps.
        }
\label{fig:ps_img}
\end{figure*}

\subsection{Source detection in NIKA2 data}
NIKA2 sources are detected iteratively, at the positions of S/N peaks with a threshold of 4$\sigma$, using the following procedure (see \citealt{Ricci2018PhD} for further details).
1) The maps are filtered as $S = \left(G_{\theta_1} \ast M - G_{\theta_2} \ast M \right) / N$, where $G_{\theta}$ is a Gaussian filter with FWHM $\theta$ and $M$ is the NIKA2 150 or 260 GHz map. We use the respective beam FWHM for $\theta_1$ and $\theta_2$ is set to 75 arcsec. This allows us to amplify the signal from point sources by removing the noise below the telescope resolution, and large scale atmospheric residual noise fluctuations. The signal is normalized by the standard deviation map $N$ so that it is expressed in units of S/N.
2) A source is fitted on the map $M$ using a Gaussian function corresponding to the NIKA2 beam plus a local background, at the location of the highest signal to noise ratio, but the precise location is allowed to vary within one beam FWHM. 
3) The best-fit source model is subtracted from the NIKA2 map $M$.
4) We repeat steps 1, 2, and 3 until no source is detected above the chosen signal to noise ratio threshold on the map $S$.

In the end, we obtain a point source catalog with fluxes and coordinates, as well as a point source model map. We note that the fluxes are corrected from transfer function filtering effects, which are estimated by injecting and recovering point sources in processed data (filtering factor of about 15\%). The correlated noise is accounted for in the uncertainties using MC noise simulations. The catalog purity is estimated by applying the same procedure in half-difference maps. Given the S/N threshold and the detection parameters, it is estimated to be 0.88 at 150 GHz and 0.93 at 260 GHz. Once the source catalogs are made in each band independently, we measure the flux of the counterpart band by fitting a source at the location of the detection. We match the catalogs with other bands to associate the detected sources using an aperture of one beam FWHM. We also list the possible matches within the same band of two nearby sources if they fall within a single beam.

The list of NIKA2 identified sources in both bands, within 5 arcmin of the cluster centers, are reported in Table~\ref{tab:ps_nika2} for all clusters. In Figure~\ref{fig:ps_img}, the source positions are reported on the images.

\begin{table*}[h]
\caption{\footnotesize{Point sources detected with NIKA2 in the fields of XLSSC~072, XLSSC~100, and XLSSC~102.}}
\begin{center}
\resizebox{\textwidth}{!} {
\begin{tabular}{lrrrrrrrrrrrr}
\toprule
Name & Label &   S/N &       R.A. &     Dec. &   Distance$^\dagger$ &  $F_{\rm detection}$ &  $\Delta F_{\rm detection}$ & $F_{\rm counterpart}$ &  $\Delta F_{\rm counterpart}$ & Match \\
   &    &    &       [deg] &     [deg] &   [arcsec] &   [mJy] &  [mJy] &  [mJy] &  [mJy] & \\
\midrule
\multicolumn{11}{l}{Field of XLSSC 072} \\
\hline
\hline
\multicolumn{11}{c}{Detection at 260 GHz, counterparts at 150 GHz} \\
\hline
 NIKA2-260 J021511.6-034309 &   N260-1 &  25.5 &  33.7982 & -3.7192 &  187.8 &  20.48 &      0.79 &   30.15 &        0.15 &          N260-10, N150-1, RS-1 \\
 NIKA2-260 J021533.8-034025 &   N260-2 &  14.9 &  33.8909 & -3.6738 &  238.5 &  12.50 &      0.82 &    2.86 &        0.18 &                        N150-2,  RS-14 \\
 NIKA2-260 J021527.0-034203 &   N260-3 &   9.9 &  33.8623 & -3.7009 &  100.5 &   5.75 &      0.56 &    0.95 &        0.12 &  N260-6, N260-15, N150-3       \\
 NIKA2-260 J021516.4-034300 &   N260-4 &   8.1 &  33.8182 & -3.7167 &  118.9 &   5.27 &      0.62 &    0.40 &        0.13 &                       \\
 NIKA2-260 J021537.0-034354 &   N260-5 &   6.5 &  33.9040 & -3.7318 &  195.0 &   4.43 &      0.68 &    0.15 &        0.15 &             \\
 NIKA2-260 J021526.4-034159 &   N260-6 &   5.9 &  33.8600 & -3.6998 &  100.9 &   4.91 &      0.56 &    0.72 &        0.12 &           N260-3        \\
 NIKA2-260 J021535.3-034024 &   N260-7 &   5.8 &  33.8969 & -3.6735 &  253.3 &   4.88 &      0.87 &    0.63 &        0.19 &                \\
 NIKA2-260 J021527.4-034129 &   N260-8 &   5.6 &  33.8641 & -3.6914 &  134.4 &   3.12 &      0.60 &    0.12 &        0.13 &                \\
 NIKA2-260 J021531.6-034041 &   N260-9 &   5.3 &  33.8818 & -3.6781 &  207.0 &   3.57 &      0.73 &    0.20 &        0.16 &           \\
 NIKA2-260 J021512.2-034309 &  N260-10 &   5.0 &  33.8010 & -3.7192 &  177.7 &   9.42 &      0.76 &   16.06 &        0.15 &           N260-1,  N150-1   RS-1 \\
 NIKA2-260 J021528.6-034511 &  N260-11 &   4.9 &  33.8690 & -3.7531 &  119.1 &   2.93 &      0.57 &    0.41 &        0.13 &                \\
 NIKA2-260 J021536.5-034526 &  N260-12 &   4.5 &  33.9022 & -3.7573 &  218.9 &   3.46 &      0.74 &    0.35 &        0.17 &              \\
 NIKA2-260 J021528.5-034250 &  N260-13 &   4.4 &  33.8688 & -3.7139 &   80.5 &   2.28 &      0.53 &    0.66 &        0.11 &              N150-4, N150-6        \\
 NIKA2-260 J021533.8-034247 &  N260-14 &   4.3 &  33.8906 & -3.7131 &  153.2 &   2.70 &      0.61 &    0.39 &        0.13 &             \\
 NIKA2-260 J021527.1-034211 &  N260-15 &   4.3 &  33.8628 & -3.7032 &   94.0 &   4.00 &      0.55 &    0.93 &        0.12 &           N260-3, N150-3         \\
 NIKA2-260 J021531.8-034425 &  N260-16 &   4.2 &  33.8826 & -3.7404 &  128.2 &   2.45 &      0.57 &    0.46 &        0.13 &        \\
 NIKA2-260 J021533.5-034320 &  N260-17 &   4.2 &  33.8894 & -3.7225 &  142.3 &   2.51 &      0.59 &    0.38 &        0.13 &        \\
 NIKA2-260 J021531.8-034509 &  N260-18 &   4.2 &  33.8824 & -3.7527 &  151.2 &   2.33 &      0.61 &    0.09 &        0.14 &                  \\
 NIKA2-260 J021539.3-034205 &  N260-19 &   4.1 &  33.9136 & -3.7015 &  244.9 &   3.53 &      0.82 &    1.16 &        0.18 &                     N150-5         \\
 NIKA2-260 J021505.7-034157 &  N260-20 &   4.1 &  33.7739 & -3.6992 &  290.1 &   5.72 &      1.36 &    0.03 &        0.25 &                    \\
 NIKA2-260 J021523.1-034613 &  N260-21 &   4.0 &  33.8462 & -3.7703 &  160.2 &   2.80 &      0.65 &   -0.15 &        0.14 &                \\
 NIKA2-260 J021529.6-034300 &  N260-22 &   4.0 &  33.8732 & -3.7168 &   89.8 &   2.09 &      0.53 &    0.50 &        0.12 &                    \\
 NIKA2-260 J021513.9-034417 &  N260-23 &   4.0 &  33.8077 & -3.7382 &  158.1 &   2.75 &      0.70 &    0.53 &        0.14 &                    \\
\hline
\multicolumn{11}{c}{Detection at 150 GHz, counterparts at 260 GHz} \\
\hline
 NIKA2-150 J021511.5-034308 &  N150-1 &  199.0 &  33.7979 & -3.7190 &  188.8 &  30.30 &      0.15 &   19.71 &        0.81 &          N260-1, N260-10,   RS-1 \\
 NIKA2-150 J021533.8-034025 &  N150-2 &   16.6 &  33.8910 & -3.6739 &  238.5 &   2.84 &      0.17 &   12.47 &        0.84 &               N260-2,  RS-14 \\
 NIKA2-150 J021527.0-034207 &  N150-3 &    8.6 &  33.8626 & -3.7021 &   97.1 &   1.12 &      0.11 &    4.97 &        0.57 &      N260-3, N260-6, N260-15        \\
 NIKA2-150 J021528.3-034248 &  N150-4 &    6.5 &  33.8681 & -3.7133 &   79.5 &   0.81 &      0.11 &    2.06 &        0.54 &  N150-6,           N260-13        \\
 NIKA2-150 J021539.2-034205 &  N150-5 &    6.2 &  33.9134 & -3.7016 &  244.1 &   1.01 &      0.18 &    3.43 &        0.84 &                  N260-19        \\
 NIKA2-150 J021528.5-034300 &  N150-6 &    4.5 &  33.8686 & -3.7168 &   74.7 &   0.74 &      0.11 &    1.40 &        0.54 &  N150-4,         N260-13, N260-22        \\
 NIKA2-150 J021522.7-034241 &  N150-7 &    4.2 &  33.8446 & -3.7115 &   55.6 &   0.43 &      0.11 &    0.18 &        0.55 &          \\
 \bottomrule
\multicolumn{11}{l}{Field of XLSSC 100} \\
\hline
\hline
\multicolumn{11}{c}{Detection at 260 GHz, counterparts at 150 GHz} \\
\hline
 NIKA2-260 J020618.6-061123 &   N260-1 &  7.5 &  31.5774 & -6.1898 &  102.5 &  4.46 &      0.57 &    0.60 &        0.13 &      N150-5        \\
 NIKA2-260 J020612.4-061038 &   N260-2 &  6.4 &  31.5515 & -6.1774 &   56.7 &  3.62 &      0.55 &    0.97 &        0.12 &         N150-1        \\
 NIKA2-260 J020606.8-061357 &   N260-3 &  5.6 &  31.5284 & -6.2327 &  160.6 &  4.09 &      0.71 &    0.84 &        0.16 &        N150-2        \\
 NIKA2-260 J020622.2-060801 &   N260-4 &  5.3 &  31.5925 & -6.1337 &  264.1 &  5.12 &      0.97 &    0.74 &        0.22 &         N150-6       \\
 NIKA2-260 J020616.1-061012 &   N260-5 &  4.7 &  31.5669 & -6.1701 &  104.4 &  2.74 &      0.58 &    0.41 &        0.13 &                       \\
 NIKA2-260 J020618.4-061217 &   N260-6 &  4.6 &  31.5766 & -6.2049 &  107.8 &  2.69 &      0.58 &    0.22 &        0.13 &                       \\
 NIKA2-260 J020613.8-061313 &   N260-7 &  4.6 &  31.5574 & -6.2204 &  103.3 &  2.78 &      0.59 &    0.23 &        0.13 &                       \\
 NIKA2-260 J020627.6-061008 &   N260-8 &  4.3 &  31.6150 & -6.1691 &  251.4 &  3.82 &      0.89 &    0.89 &        0.21 &         N150-3        \\
 NIKA2-260 J020628.0-061344 &   N260-9 &  4.2 &  31.6165 & -6.2290 &  274.3 &  4.36 &      1.00 &    0.25 &        0.24 &                       \\
 NIKA2-260 J020609.4-061306 &  N260-10 &  4.1 &  31.5392 & -6.2184 &   97.8 &  2.65 &      0.60 &   -0.05 &        0.13 &                       \\
 NIKA2-260 J020602.6-061204 &  N260-11 &  4.1 &  31.5109 & -6.2013 &  139.5 &  2.93 &      0.70 &   -0.02 &        0.15 &                      \\
 \hline
\multicolumn{11}{c}{Detection at 150 GHz, counterparts at 260 GHz} \\
\hline
 NIKA2-150 J020612.4-061040 &  N150-1 &  7.8 &  31.5516 & -6.1778 &   55.6 &  0.93 &      0.12 &    3.64 &        0.57 &         N260-2        \\
 NIKA2-150 J020606.7-061357 &  N150-2 &  5.3 &  31.5281 & -6.2327 &  161.4 &  0.88 &      0.15 &    4.06 &        0.73 &         N260-3        \\
 NIKA2-150 J020627.4-061011 &  N150-3 &  4.8 &  31.6140 & -6.1697 &  247.3 &  1.04 &      0.20 &    3.23 &        0.90 &         N260-8        \\
 NIKA2-150 J020624.9-061056 &  N150-4 &  4.8 &  31.6039 & -6.1824 &  200.3 &  0.83 &      0.17 &    2.11 &        0.75 &                    \\
 NIKA2-150 J020618.6-061123 &  N150-5 &  4.4 &  31.5775 & -6.1898 &  102.6 &  0.56 &      0.13 &    4.41 &        0.59 &         N260-1        \\
 NIKA2-150 J020622.5-060800 &  N150-6 &  4.0 &  31.5937 & -6.1334 &  267.7 &  0.91 &      0.22 &    4.55 &        1.02 &         N260-4        \\
 \bottomrule
\multicolumn{11}{l}{Field of XLSSC 102} \\
\hline
\hline
\multicolumn{11}{c}{Detection at 260 GHz, counterparts at 150 GHz} \\
\hline
 NIKA2-260 J020517.7-044037 &  N260-1 &  5.6 &  31.3239 & -4.6770 &   90.3 &  3.40 &      0.61 &    0.38 &        0.15 &         N150-4        \\
 NIKA2-260 J020459.7-043852 &  N260-2 &  4.9 &  31.2486 & -4.6478 &  263.7 &  6.07 &      1.15 &    0.92 &        0.27 &                       \\
 NIKA2-260 J020522.4-043903 &  N260-3 &  4.7 &  31.3434 & -4.6509 &   76.9 &  2.84 &      0.59 &    0.61 &        0.14 &         N150-2        \\
 NIKA2-260 J020502.6-043824 &  N260-4 &  4.4 &  31.2607 & -4.6401 &  224.2 &  4.31 &      0.93 &    0.41 &        0.22 &                       \\
 NIKA2-260 J020509.2-044100 &  N260-5 &  4.3 &  31.2881 & -4.6834 &  165.9 &  3.32 &      0.75 &    0.30 &        0.18 &                       \\
 NIKA2-260 J020520.6-043843 &  N260-6 &  4.2 &  31.3357 & -4.6454 &   54.5 &  2.50 &      0.57 &    0.34 &        0.14 &                  RS-4 \\
\hline
\multicolumn{11}{c}{Detection at 150 GHz, counterparts at 260 GHz} \\
\hline
 NIKA2-150 J020511.8-044305 &  N150-1 &  4.7 &  31.2991 & -4.7182 &  252.2 &  1.20 &      0.25 &    2.95 &        1.05 &                       \\
 NIKA2-150 J020522.3-043905 &  N150-2 &  4.6 &  31.3429 & -4.6516 &   75.2 &  0.63 &      0.14 &    2.67 &        0.59 &         N260-3        \\
 NIKA2-150 J020536.1-043748 &  N150-3 &  4.4 &  31.4006 & -4.6301 &  292.9 &  1.46 &      0.30 &    5.05 &        1.30 &                       \\
 NIKA2-150 J020517.6-044047 &  N150-4 &  4.2 &  31.3235 & -4.6798 &  100.3 &  0.56 &      0.15 &    2.36 &        0.62 &         N260-1        \\
 \bottomrule
\end{tabular}
}
\end{center}
        {\small {\bf Notes.} 
        $^\dagger$ distance from the cluster reference center.}
\label{tab:ps_nika2}
\end{table*}

\subsection{Submillimeter contamination}
We compute the mean 150 GHz to 260 GHz flux ratio for all the sources detected at 150 GHz, excluding radio sources, $F_{150}/F_{260} = 0.221 \pm 0.014$. Using this reference value, we find that the expected S/N is nearly the same at 150 and 260 GHz, implying that any source that could significantly bias the SZ signal should be detected at 260 GHz. If instead we use the mean ratio for the sources detected at 260 GHz, $F_{150}/F_{260} = 0.123 \pm 0.008$, the S/N should be two times larger at 260 GHz, and the potential bias would be reduced. Accordingly, and given the fact that no hint of a 260 GHz source is visible in the SZ region for the three clusters, we do not expect that any significant submillimeter source that is blended in the SZ signal could be missed and significantly bias the clusters analysis.

\begin{figure}
        \centering
        \includegraphics[width=0.45\textwidth]{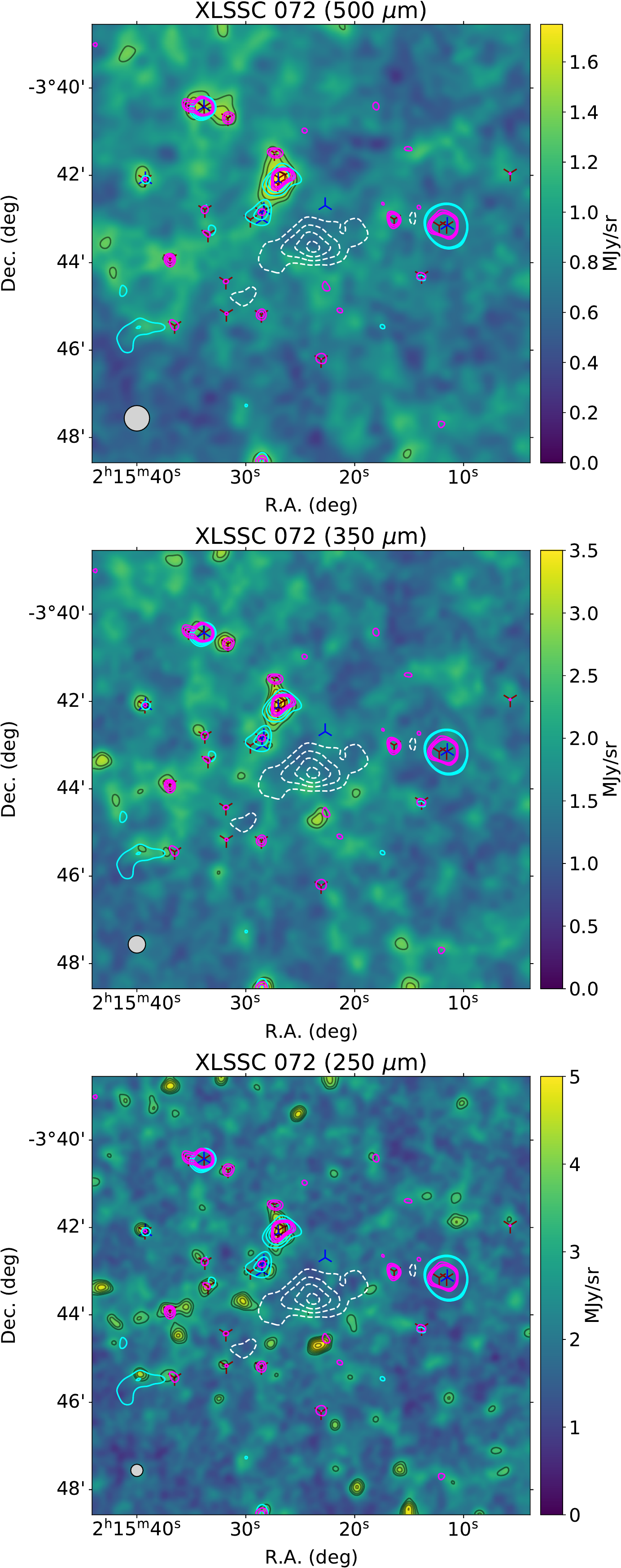}
        \caption{Herschel/PACS images of XLSSC~072 at 500, 350, and 250 $\mu$m (from top to bottom). Black contours give the S/N in units of 1$\sigma$, starting at 3$\sigma$. The gray circles in the bottom left corner show the PACS beam FWHM in each band. White dashed contours show the NIKA2 150 GHz contours at -3, -5, -7, and -9 $\sigma$. Magenta and cyan contours show the 3, 4, and 5$\sigma$ NIKA2 S/N at 150 and 260 GHz, respectively. The blue and red crosses indicate the point sources identified at 150 and 260 GHz in the NIKA2 data.}
\label{fig:herschel}
\end{figure}

The field of XLSSC~072 was observed with Herschel/SPIRE \citep{Griffin2010} at 500, 350, and 250$\mu$m (obsID 1342189031 and 1342190313), and we compare the SPIRE maps to NIKA2 images in Figure~\ref{fig:herschel}. Although the image depth is relatively shallow, the brightest regions match the NIKA2 sources well. This confirms that no bright submillimeter source is missed by NIKA2 and that the SZ signal from XLSSC~072 is not contaminated.

\subsection{Radio GMRT counterparts}
While NIKA2 260 GHz data can be used to assess the contamination from submillimeter sources, radio data are necessary to address the contamination from radio galaxies. We use XXL/GMRT images and catalogs from \cite{Smolcic2018}, hereafter XXL Paper XXIX, to do so. In Table~\ref{tab:radio_table}, we list all the GMRT sources identified in the 10 arcmin $\times$ 10 arcmin region around the three clusters. The GMRT positions uncertainties are at most 0.5 arcsec, which is negligible for our purpose. In addition, we provide estimates of the fluxes expected at 150 GHz assuming a power-law spectrum: $F_{\nu} = F_0 \left(\frac{\nu}{\nu_0}\right)^{\alpha}$. This is done by using a spectral index equal to the mean value of the sample selected in Paper XXIX, between 610 and 1400 MHz (or its mean value plus one standard deviation). This provides an upper limit of the expected flux since a steepening of the radio spectrum is common at higher frequencies. Some of the GMRT sources are also detected in the NVSS surveys, in which case a spectral index estimate is available for individual sources, which we use to compute a more reliable flux estimate at 150 GHz. Given these estimates, only XXL-GMRT~J021511.4-034309 in the field of XLSSC~072 was expected to be detected and is indeed detected (NIKA2-150~J021511.5-034308, NIKA2-260~J021511.6-034309). Figure~\ref{fig:ps_img} shows the GMRT maps and how they compare to the NIKA2 data. We also use the FIRST catalog, which we compare to the GMRT images. We note that it is very unlikely that radio sources that are not detected in the NVSS or FIRST significantly affect NIKA2 data given the survey sensitivity (down to 0.45 and 0.15 mJy/beam). No radio source is expected to significantly contaminate the SZ signal, at least in the region where the two could be blended.

\begin{table*}[h]
\caption{\footnotesize{Radio sources identified around the three clusters with GMRT.}}
\begin{center}
\resizebox{\textwidth}{!} {
\begin{tabular}{llrrrrrrrrrr}\toprule
Name & Label &   S/N &       R.A. &     Dec. &   Distance &  $F_{\rm 610 \ MHz}$ &  $\Delta F_{\rm 610 \ MHz}$ &  $\alpha_{\rm 610-1400 \ MHz}$ &  $F_{\rm 150 \ GHz} (\alpha_{\rm mean})$ &  $F_{\rm 150 \ GHz} (\alpha_{\rm mean}+\sigma_{\alpha})$ & $F_{\rm 150 \ GHz} (\alpha_{\rm 610-1400 \ MHz})$  \\
   &    &    &       [deg] &     [deg] &   [arcsec] &   [mJy] &  [mJy] & & [mJy] &  [mJy] & [mJy] \\
\midrule
  \multicolumn{12}{l}{Field of XLSSC 072} \\
\hline
\hline
 XXL-GMRT J021511.4-034309 &   RS-1 &  2628.3 &  33.7979 & -3.7192 &  188.8 &  313.29 &      0.23 &   0.60 &         5.05 &          32.79 &         11.52 \\
 XXL-GMRT J021536.1-034423 &   RS-2 &   357.6 &  33.9008 & -3.7399 &  189.2 &   28.52 &      0.12 &   0.96 &         0.46 &           2.99 &          0.14 \\
 XXL-GMRT J021522.5-034441 &   RS-3 &   145.9 &  33.8440 & -3.7447 &   70.7 &   11.74 &      0.08 &   0.98 &         0.19 &           1.23 &          0.05 \\
 XXL-GMRT J021516.7-034555 &   RS-4 &    37.0 &  33.8197 & -3.7654 &  178.8 &    2.94 &      0.08 &     &         0.05 &           0.31 &            \\
 XXL-GMRT J021519.5-034756 &   RS-5 &    31.2 &  33.8313 & -3.7991 &  271.6 &    2.39 &      0.08 &     &         0.04 &           0.25 &            \\
 XXL-GMRT J021534.2-034002 &   RS-6 &    28.0 &  33.8925 & -3.6673 &  260.7 &    3.87 &      0.13 &     &         0.06 &           0.40 &            \\
 XXL-GMRT J021509.3-034357 &   RS-7 &    27.3 &  33.7890 & -3.7326 &  220.4 &    2.31 &      0.08 &     &         0.04 &           0.24 &            \\
 XXL-GMRT J021530.4-034041 &   RS-8 &    26.2 &  33.8768 & -3.6781 &  197.5 &    2.23 &      0.08 &     &         0.04 &           0.23 &            \\
 XXL-GMRT J021541.7-033931 &   RS-9 &    18.0 &  33.9239 & -3.6587 &  359.4 &    1.64 &      0.09 &     &         0.03 &           0.17 &            \\
 XXL-GMRT J021506.8-034205 &  RS-10 &     9.9 &  33.7786 & -3.7014 &  271.4 &    0.86 &      0.09 &     &         0.01 &           0.09 &            \\
 XXL-GMRT J021529.2-034713 &  RS-11 &     9.5 &  33.8720 & -3.7872 &  234.1 &    0.68 &      0.07 &     &         0.01 &           0.07 &            \\
 XXL-GMRT J021540.8-034418 &  RS-12 &     9.4 &  33.9202 & -3.7383 &  256.0 &    0.75 &      0.08 &     &         0.01 &           0.08 &            \\
 XXL-GMRT J021531.9-033839 &  RS-13 &     9.0 &  33.8830 & -3.6442 &  317.4 &    0.78 &      0.09 &     &         0.01 &           0.08 &            \\
 XXL-GMRT J021533.7-034025 &  RS-14 &     8.9 &  33.8906 & -3.6737 &  238.2 &    0.71 &      0.08 &     &         0.01 &           0.07 &            \\
 XXL-GMRT J021530.2-034341 &  RS-15 &     8.5 &  33.8759 & -3.7282 &   93.4 &    0.55 &      0.06 &     &         0.01 &           0.06 &            \\
 \bottomrule
  \multicolumn{12}{l}{Field of XLSSC 100} \\
\hline
\hline
  XXL-GMRT J020619.5-061147 &   RS-1 &  115.7 &  31.5816 & -6.1966 &  117.4 &  8.14 &      0.07 &   0.83 &         0.13 &           0.85 &          0.08 \\
 XXL-GMRT J020611.3-061616 &   RS-2 &   51.3 &  31.5474 & -6.2712 &  281.6 &  3.36 &      0.07 &     &         0.05 &           0.35 &            \\
 XXL-GMRT J020611.1-061102 &   RS-3 &   22.5 &  31.5465 & -6.1840 &   33.6 &  1.71 &      0.08 &     &         0.03 &           0.18 &            \\
 XXL-GMRT J020555.8-060920 &   RS-4 &   14.6 &  31.4828 & -6.1558 &  272.2 &  1.00 &      0.07 &     &         0.02 &           0.10 &            \\
 XXL-GMRT J020620.7-061629 &   RS-5 &   13.2 &  31.5866 & -6.2749 &  324.1 &  0.77 &      0.06 &     &         0.01 &           0.08 &            \\
 XXL-GMRT J020613.9-061001 &   RS-6 &   11.9 &  31.5580 & -6.1671 &   98.6 &  0.90 &      0.08 &     &         0.01 &           0.09 &            \\
 XXL-GMRT J020618.3-060742 &   RS-7 &   10.4 &  31.5766 & -6.1285 &  252.3 &  0.83 &      0.08 &     &         0.01 &           0.09 &            \\
 XXL-GMRT J020615.6-060940 &   RS-8 &    8.3 &  31.5653 & -6.1612 &  128.5 &  0.56 &      0.07 &     &         0.01 &           0.06 &            \\
 XXL-GMRT J020606.6-061303 &   RS-9 &    8.3 &  31.5278 & -6.2178 &  117.2 &  0.48 &      0.06 &     &         0.01 &           0.05 &            \\
 XXL-GMRT J020614.8-061556 &  RS-10 &    8.1 &  31.5620 & -6.2658 &  266.2 &  0.48 &      0.06 &     &         0.01 &           0.05 &            \\
 XXL-GMRT J020602.0-061139 &  RS-11 &    7.5 &  31.5084 & -6.1942 &  145.4 &  0.45 &      0.06 &     &         0.01 &           0.05 &            \\
 \bottomrule
 \multicolumn{12}{l}{Field of XLSSC 102} \\
\hline
\hline
  XXL-GMRT J020537.2-043904 &  RS-1 &  228.3 &  31.4054 & -4.6511 &  299.3 &  12.86 &      0.09 &    0.8 &         0.21 &           1.35 &          0.16 \\
 XXL-GMRT J020523.5-043441 &  RS-2 &   71.9 &  31.3480 & -4.5781 &  281.9 &   3.34 &      0.05 &     &         0.05 &           0.35 &            \\
 XXL-GMRT J020528.7-044132 &  RS-3 &   30.3 &  31.3696 & -4.6924 &  224.3 &   1.60 &      0.05 &     &         0.03 &           0.17 &            \\
 XXL-GMRT J020520.6-043843 &  RS-4 &   24.2 &  31.3360 & -4.6454 &   55.6 &   1.24 &      0.05 &     &         0.02 &           0.13 &            \\
 XXL-GMRT J020457.2-044308 &  RS-5 &   13.3 &  31.2386 & -4.7191 &  384.6 &   0.68 &      0.05 &     &         0.01 &           0.07 &            \\
 XXL-GMRT J020520.6-044118 &  RS-6 &    8.8 &  31.3361 & -4.6886 &  141.1 &   0.47 &      0.05 &     &         0.01 &           0.05 &            \\
 XXL-GMRT J020523.6-044028 &  RS-7 &    7.8 &  31.3485 & -4.6746 &  125.1 &   0.42 &      0.05 &     &         0.01 &           0.04 &            \\
 XXL-GMRT J020520.5-044031 &  RS-8 &    7.6 &  31.3356 & -4.6755 &   97.7 &   2.18 &      0.05 &     &         0.04 &           0.23 &            \\
 XXL-GMRT J020519.8-044006 &  RS-9 &    7.3 &  31.3327 & -4.6683 &   70.1 &   2.15 &      0.05 &     &         0.03 &           0.22 &            \\
\bottomrule
\end{tabular}
}
\end{center}
        {\small {\bf Notes.} 
        The parameters $\alpha_{\rm mean} = -0.75$ and $\sigma_{\alpha}=0.34$ are the mean and standard deviation of the spectral indices measured by matching GMRT 610 MHz data and NVSS 1400 MHz data in the XXL survey (Paper XXIX). The parameter $\alpha_{\rm 610-1400 MHz}$ is the measured spectral index for the given source when detected in the NVSS \citep{Condon1998}.}
\label{tab:radio_table}
\end{table*}

\subsection{Point source model and impact on the SZ signal}
\begin{figure*}
        \centering
        \includegraphics[width=0.95\textwidth]{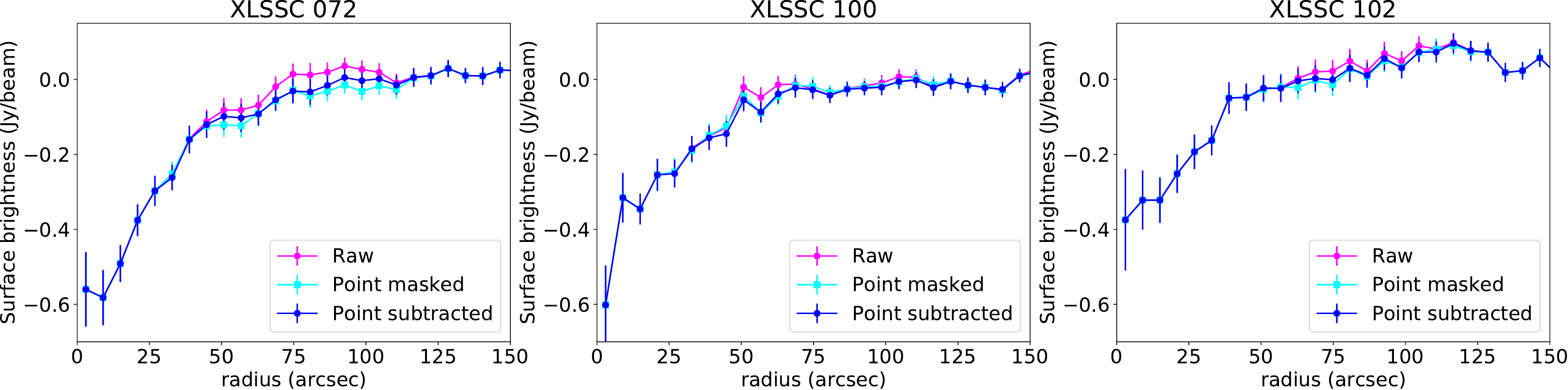}
        \caption{Comparison of 150 GHz surface brightness profiles when accounting (or not) for the point sources.}
\label{fig:ps_check_profile}
\end{figure*}

No radio or submillimeter source is expected to bias significantly the SZ signal observed with NIKA2 in XLSSC~072, XLSSC~100, or XLSSC~102, as no source is expected to be blended in the SZ signal. Nonetheless, several sources are detected within the cluster region, where the SZ signal, although fainter, extends. We built a point source model according to the 150 GHz catalog and use it to account for the point source in the SZ analysis. Since the sources are fitted using a local background, this assumes that the SZ signal is smooth at the location of the point sources. In Figure~\ref{fig:ps_check_profile}, we show the surface brightness profiles when accounting (or not) for the detected point sources. This is done either by masking the point sources or correcting for them. The point sources have a mild contribution, so uncertainties in the model should be negligible.

\section{Thermal electron density profiles}\label{app:density_profiles}
\begin{figure*}
        \centering
        \includegraphics[width=0.99\textwidth]{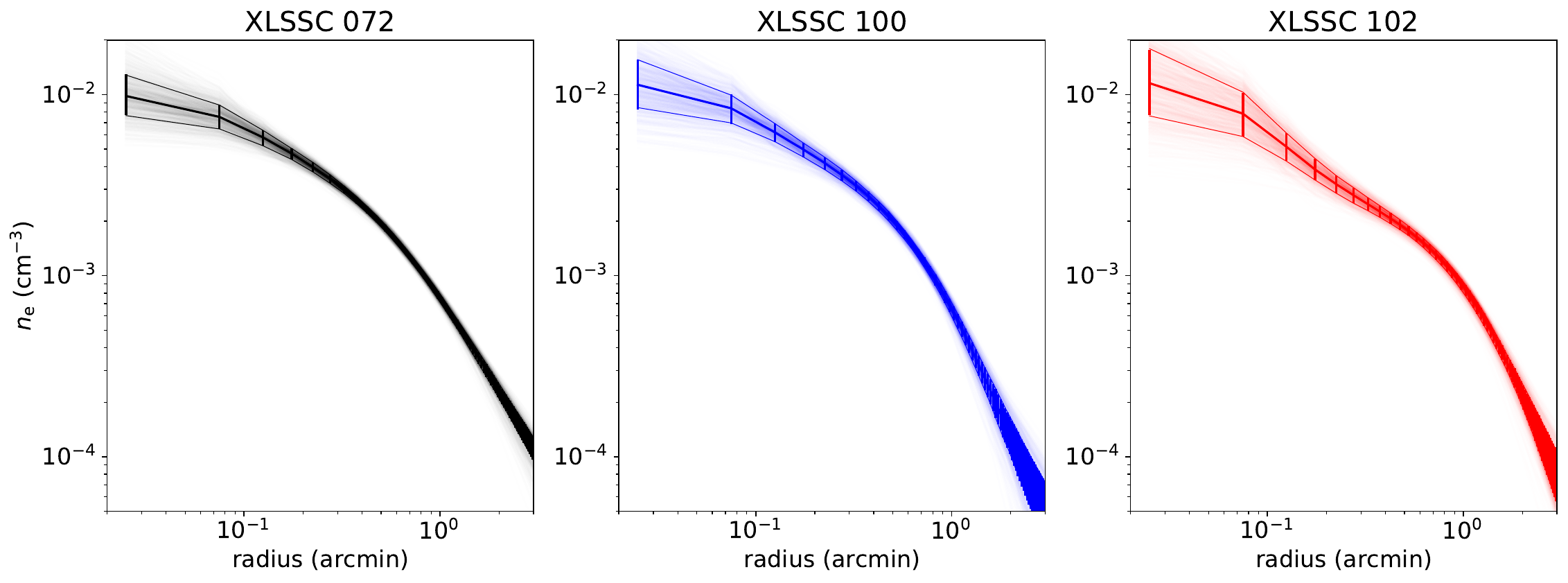}
        \caption{Thermal electron density profile, from left to right, of XLSSC~072, XLSSC~100, and XLSSC~102. The solid lines provide the best profiles and the 68\% confidence interval. The 1000 MC realizations are also provided via transparent markings to show the dispersion.}
\label{fig:density_profile}
\end{figure*}

The electron density profiles of the three clusters derived from XMM-Newton data are reported in Figure~\ref{fig:density_profile}.

\section{Thermodynamic profile diagnosis}\label{app:thermo_profile_diagnosis}
This appendix presents the temperature, entropy, and gas fraction profiles of our target clusters, which we use as a thermodynamic diagnosis of the dynamical state and to address systematic effects in the mass measurement. They are computed within the framework of the \textsc{MINOT} software \citep{Adam2020} given the pressure and the density inferred from the SZ and X-ray data. As a reference, we use the pressure profile inferred from the gNFW model fit to the data. We note that detailed discussions regarding the recovered temperature, entropy and gas fraction were presented in Paper XLIV for XLSSC~102. The thermodynamic profiles reported here are in agreement with our previous analysis, but they differ slightly due to the updated density profile.

\begin{figure*}
        \centering
        \includegraphics[width=0.33\textwidth]{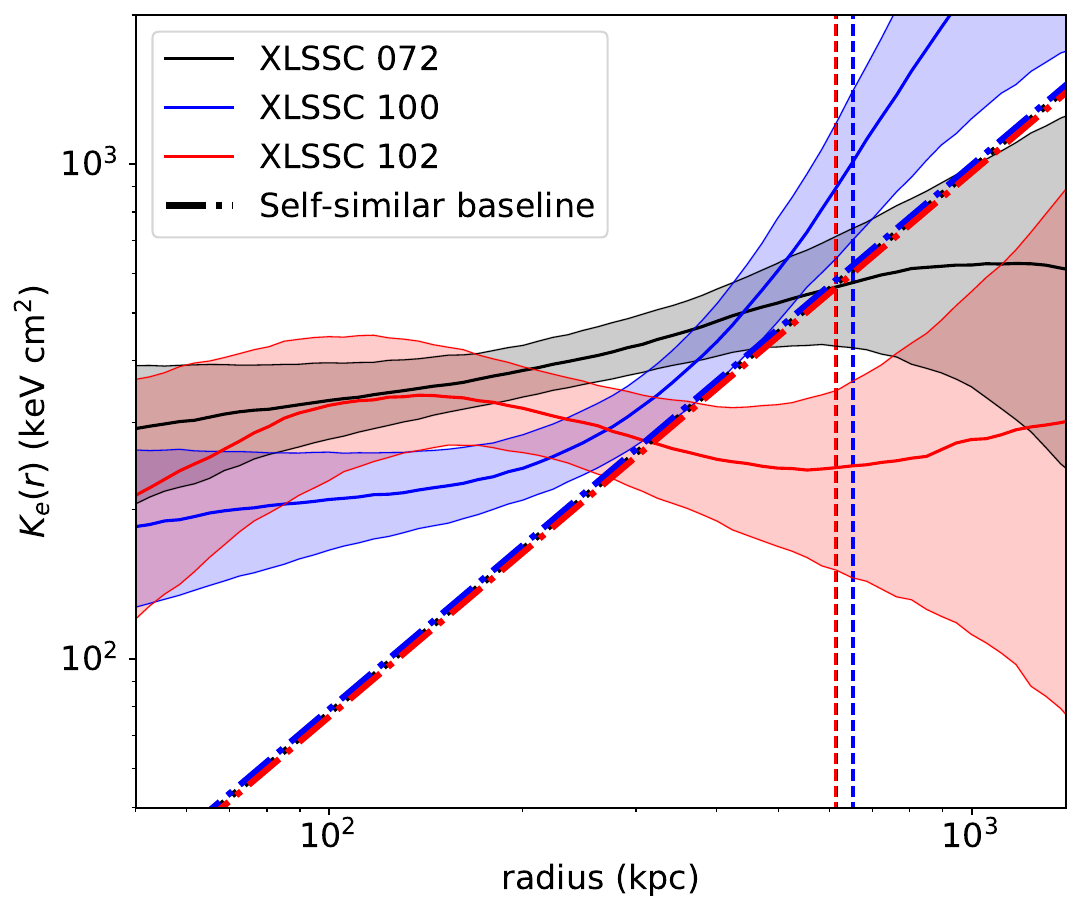}
        \includegraphics[width=0.33\textwidth]{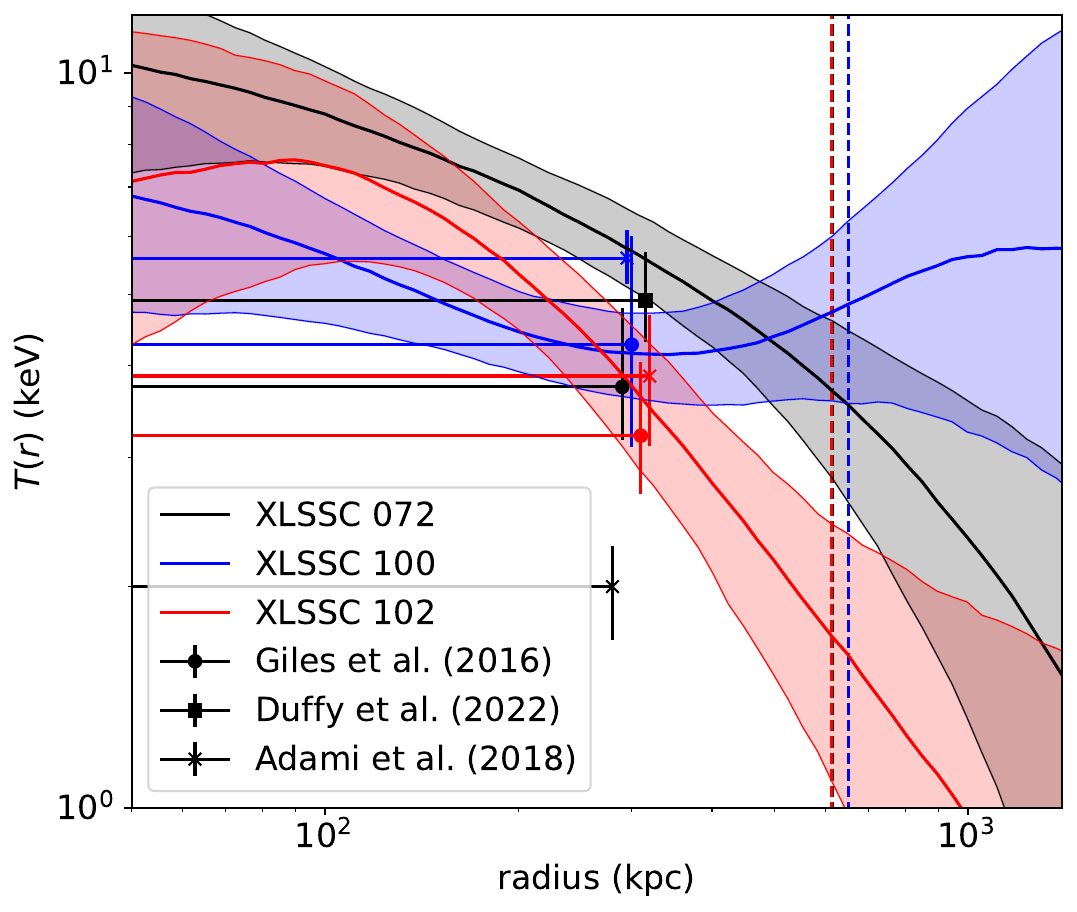}
        \includegraphics[width=0.33\textwidth]{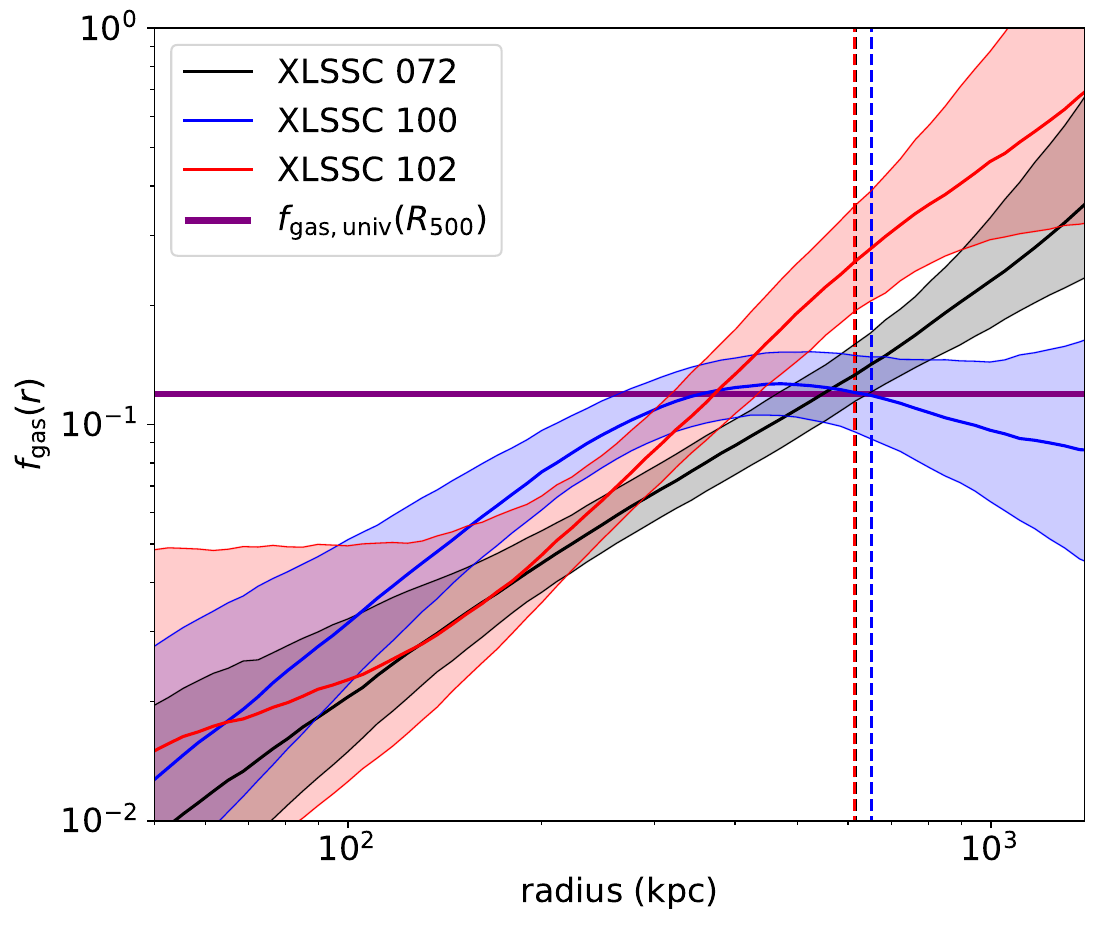}
        \caption{
        Thermodynamic profiles. 
        {\bf Left:} Entropy profiles derived by combining X-ray and SZ measurement. The self-similar baseline, accounting only for gravitational effects \citep{Voit2005b}, is reported given the masses derived via the $Y_{\rm X} - M$ relation (as well as $R_{500}$ shown by the vertical dashed lines).
        {\bf Middle:} Temperature profiles derived by combining X-ray and SZ measurement. X-ray spectroscopic measurement from Paper III and Paper XX, obtained within 300 kpc, are reported. For XLSSC~072, we also report the result from Paper XLVIII, also obtained with 300 kpc from X-ray spectroscopy.
        {\bf Right:} Gas fraction profiles derived by combining X-ray and SZ measurement, assuming $b_{\rm HSE} = 0$. The mean cosmic value from \cite{PlkanckXIII2016} is reported for reference.
        }
\label{fig:thermo_profile}
\end{figure*}

The entropy and temperature are given by
\begin{equation}
K_{\rm e}(r) = \frac{P_{\rm e}(r)}{n_{\rm e}(r)^{5/3}}
\end{equation}
and 
\begin{equation}
k_{\rm B} T(r)= \frac{P_{\rm e}(r)}{n_{\rm e}(r)},
\end{equation}
respectively. The gas fraction is computed as
\begin{equation}
f_{\rm gas}(r) = \frac{M_{\rm gas}(r)}{M_{\rm tot}(r)},
\label{eq:fgas}
\end{equation}
where the gas mass is obtained by integrating the electron density profile as
\begin{equation}
M_{\rm gas}(r) = 4 \pi \int_0^r \mu_{\rm e} m_{\rm p} n_{\rm e}(r^{\prime}) {r^{\prime}}^2 dr^{\prime},
\label{eq:gas_mass_profile}
\end{equation}
with the mean molecular weight $\mu_e \simeq 1.15$ and $m_p$ is the proton mass. The total mass is related to the hydrostatic mass via
\begin{equation}
M_{\rm tot}(r) = \frac{M_{\rm HSE}(r)}{\left(1-b_{\rm HSE}\right)},
\label{eq:Mtot}
\end{equation}
where $b_{\rm HSE}$ is the hydrostatic mass bias. The entropy, temperature and gas fraction profiles are shown in Figure~\ref{fig:thermo_profile} for the three clusters. 

The three entropy profiles are compared to the self-similar baseline, in the case where only gravitational effect are present \citep{Voit2005b}, given by $K(r) \propto \left(r / R_{500}\right)^{1.1}$. The masses used for the comparison are obtained from the $Y_{\rm X}$ proxy. We can observe that all clusters present a large excess entropy in the core, within $r \lesssim 300$ kpc, of about 300 keV cm$^2$. Such a feature is typical for disturbed systems \citep{Pratt2010}. This indicates that all three systems are dynamically disturbed, most likely because of the presence of merging events, in agreement with our imaging analysis (see Section~\ref{sec:dynamics}). We note that XLSSC~102 agrees with a flat entropy profile at all scales. Beyond $r \sim 300$ kpc, the profile is even lower than the self-similar baseline, although uncertainties are becoming very large. As the self-similar baseline corresponds to a minimal heat injection from gravitational collapse, such a feature is not expected. This could indicate an excess in the density profile caused by inhomogeneities in the gas, as observed for other nearby clusters with high-quality data \citep{Tchernin2016}, which would also bias low the HSE mass estimates. We note that this feature is reduced when using the direct HSE mass measurement since they are lower, but it does not entirely disappear.

The three temperature profiles agree with that of merging systems, with a profile decreasing from the core to the outskirt. We also report the projected temperature measured within 300 kpc from the core by Paper III and Paper XX. The same measurement from Paper XLVIII is also reported for XLSSC~072. Given the uncertainties and the fact that the two measurements are not strictly comparable, good qualitative agreement is observed for XLSSC~100 and XLSSC~102. On the other hand, our SZ plus X-ray-derived temperature for XLSSC~072 is in qualitative agreement with the one from Paper XLVIII, higher than but still comparable to the one from Paper III, but in significant disagreement with the one from Paper XX. As the $Y_{\rm X}$-based masses are using Paper III temperatures, we conclude that while no significant issue is observed with XLSSC~102 and XLSSC~100, the $Y_{\rm X}$-derived mass for XLSSC~072 might be biased depending on the choice of the temperature measurement. For instance, using our SZ plus X-ray measurement would increase the mass within $\sim 2 \sigma$.

The gas fraction profiles increase from the center to the outskirts, in agreement with the expected baryon depletion generally expected in the center. As we can see from Equation~\ref{eq:fgas} and \ref{eq:Mtot}, the gas fraction depends on the hydrostatic mass bias, which is set to $b_{\rm HSE} = 0$ here. On the other hand, the universal gas fraction at $R_{500}$ is given by 
\begin{equation}
f_{\rm gas, univ} (R_{500}) = \frac{\Omega_{\rm b}}{\Omega_{\rm m}} Y_{{\rm b}, 500} - f_{\star, 500},
\end{equation}
with $Y_{b, 500} \simeq 0.85$ being the baryon depletion factor and where $f_{\star, 500} \simeq 0.015$ accounts for the baryons that have condensed into stars (see \citealt{Eckert2019} for details). Assuming that the profiles should reach the universal gas fraction at $R_{500}$, it is possible to estimate the value of $b_{\rm HSE}$. As we can see, the three profiles reach the cosmic baryon fraction at about $R_{500}$. While XLSSC~072 and XLSSC~100 are in rough agreement with the expected value, XLSSC~102 presents a gas fraction that is higher by about $2 \sigma$. This might indicate that this system presents a high hydrostatic mass bias, in agreement with the fact that it is the most disturbed of our targets and already indicated thanks to the entropy profile. More quantitatively, a value of $1-b_{\rm HSE}$ of 0.85, 1.0, and 0.5 would bring XLSSC~072, XLSSC~100, and XLSSC~102 to the expected universal gas fraction value at $R_{500}$, respectively.

Given the entropy, temperature, and gas fraction profiles, we conclude that all three clusters are dynamically disturbed. Additionally, we find that the $Y_{\rm X}$-derived masses may be biased low for XLSSC~072 (within $2 \sigma$), and that direct HSE-derived masses may be biased low by up to a factor of two for XLSSC~102.

\section{SZ residuals}\label{app:residual_SZ}
\begin{figure*}
        \centering
        \includegraphics[width=0.89\textwidth]{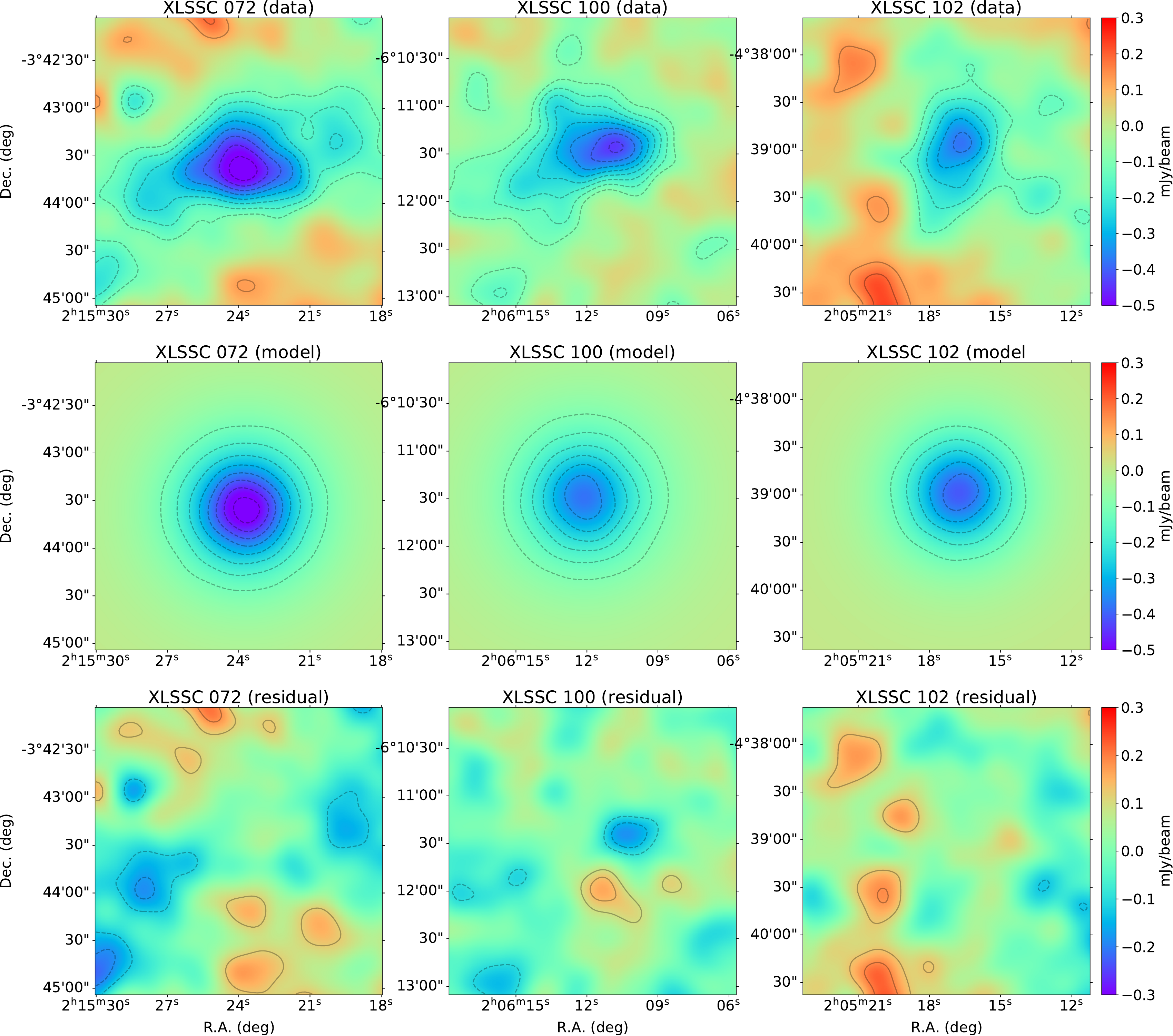}
        \caption{NIKA2 SZ maps (top), best-fit gNFW model (middle), and residual (bottom) maps for the three clusters. The contours are given in units of $2 \sigma$ starting at $\pm 2 \sigma$.}
\label{fig:SZ_residual}
\end{figure*}

The comparison between the SZ map and their best-fit gNFW model is shown in Figure~\ref{fig:SZ_residual}. While some residual structures can reach about $3 \sigma$ due to deviation from spherical symmetry, the best-fit model provides a fair description of the data in all cases.

\end{appendix}

\end{document}